\documentclass[11pt]{article}
\usepackage{verbatim}
\usepackage{a4}
\usepackage{a4wide}
\usepackage{amssymb}
\usepackage{amsmath}
\usepackage{color}
\usepackage{array}
\usepackage{lscape}
\usepackage{subeqnarray}
\usepackage{graphics,epsfig}
\usepackage{graphicx}
\usepackage{amsfonts}
\def\ie{{\rm i.e.,\/}\ }

\def\one{\mbox{\rm 1}\hskip-2.8pt \mbox{\rm l}}

\newcommand{\ZZ}{\mathbb{Z}}

\newcommand{\CC}{\mathbb{C}}

\newcommand{\nco}{\newcommand}

\nco{\munite}{\ensuremath{\,\,\mathrm{l}\!\!\!1}} 
\nco{\dps}{\displaystyle}
\nco{\ov}{\overline}
\nco{\ud}{\underline}
\nco{\NN}{\mathbb{N}}
\author{
 
 Author 1\footnotemark[1],
 Author 2\footnotemark[1],
\\  
 Author 3 on next line\footnotemark[1],
 Author 4\footnotemark[2],
}

\title{ From conformal embeddings to quantum symmetries: \\ an exceptional $SU(4)$ example \footnote{IOP, Journal of Physics, Conference Series:
International Conference on Non-Commutative Geometry and Physics,  23-27
April 2007, Laboratoire de Physique Th\'eorique, Universit\'e Paris XI,
France.  Expected online publication: First quarter 2008.}{}}

\author{

\bf{R. Coquereaux} \footnotemark[1] ${ }^{ ,}$\footnotemark[2]
$\;$and$\;$
\bf{G. Schieber} \footnotemark[1] ${ }^{ ,}$\footnotemark[3]
\\  
}

\date{}

\begin{document}
\maketitle

\footnotetext[1]{{\it Centre de Physique Th\'eorique(CPT)}, {\scriptsize {\it Luminy, Marseille}. UMR 6207 du CNRS
et des Universit\'es Aix-Marseille I, Aix-Marseille II,
et du Sud Toulon-Var,  affili\'e \`a la FRUMAM (FR 2291).}}
\footnotetext[2]{~Email: \scriptsize{Robert.Coquereaux@cpt.univ-mrs.fr}}
\footnotetext[3]{~Email: \scriptsize{schieber@cpt.univ-mrs.fr}}
\addtocounter{footnote}{0}

\abstract{}

We briefly discuss several algebraic tools that are used to describe the quantum symmetries of Boundary Conformal Field Theories on a torus. The starting point is a fusion category,  together with an action on another category described by a quantum graph.
 For known examples, the corresponding modular invariant partition function, which is sometimes associated with a conformal embedding,  provides enough information to recover the whole structure.
We illustrate these notions with the example of the conformal embedding of $SU(4)$ at level $4$ into $Spin(15)$ at level $1$, leading to the exceptional quantum graph ${\mathcal E}_4 (SU(4))$. 
 
\vspace{1.0cm}


\section*{Foreword}

There are many ways to describe the algebraic structures underlying boundary conformal field theories on a torus. Because of its concision, we choose the categorical description that was sketched in  \cite{Ostrik}, in the framework of the  $SU(2)$ classification  ($ADE$). We refer to \cite{RobertGilSL3Categories} for a more detailed presentation along those lines, in the framework of the $SU(3)$ classification. Many properties of the associated quantum graphs can also be found there, in particular the corresponding quantum groupo\"\i d and the Ocneanu algebra of quantum symmetries. The purpose of the present paper is to summarize this presentation, to show how conformal embeddings relate to this description, to present one exceptional example of type $SU(4)$, starting from a conformal embedding, and to study its algebra of quantum symmetries.

\section{Quantum graphs associated with a Lie group $G$ and module-categories}

\subsection{The fusion algebra}

We start from  ${\mathcal A}_k(G)$,  the  fusion category of an affine Lie algebra associated with a Lie group  $G$  at level $k$, or \mbox{equivalently}, the category of representations with non-zero $q$-dimension of the quantum group $G_q$ at roots of unity (set $q = exp(i \pi/\kappa)$, where the altitude $\kappa$, also called generalized Coxeter number, is $\kappa = k+ g_G$  with $g_G$ the dual Coxeter number of $G$. This monoidal category is modular and ribbon. Its Grothendieck ring comes with a special basis (corresponding to simple objects $\lambda_n$), it is usually called the fusion ring,  or the Verlinde algebra. 
Its structure constants, encoded by the fusion matrices $(N_n)^p_q$, are therefore non - negative integers and the rigidity property of the category (existence of duals)  implies, for the fusion ring,  the property  $(N_{\overline n})_{pq} = (N_n)_{qp} $, where $\overline n$ refers to the dual object $\overline \lambda_n$ (\ie the conjugate of the irreducible representation $\lambda_n$). In the case of $SU(2)$, it is a ring with one generator (self-conjugated). In the case of $SU(3)$, we have two generators, corresponding to the  fundamental irreducible representations that are conjugate to one another. In the case of $SU(4)$, we have three generators : one of them is self-conjugated and the two others are conjugated from one another.
 Multiplication by a chosen generator $\lambda_f$ is encoded by a particular fusion matrix $N_f$; it is a finite size matrix of dimension $r \times r$, with $r=k+1$ for $SU(2)$ , $r=(k+1)(k+2)/2$ for $SU(3)$, $r=(k+1)(k+2)(k+3)/3! $ for $SU(4)$, etc. Since its elements are non negative integers, it can be interpreted as the adjacency matrix of a graph, which is the  Cayley graph of multiplication by this generator, and we call it the McKay graph of the category.  Self - conjugated fundamental representations correspond to non-oriented edges of the graph (rather, they carry both orientations).
   One should certainly keep in mind the distinction between this monoidal category (with its objects and morphisms),  its Grothendieck ring, and the McKay graph,  but they will often be denoted by the same symbol,  ${\mathcal A}_k(G)$, or simply ${\mathcal A}_k$ for short, since $G$ is usually chosen once and for all.  In the $SU(2)$ case, the graph can be identified with the Coxeter - Dynkin diagram $A_r$, with $r=k+1$. In all cases, it  is a truncated Weyl chamber at level $k$ (a Weyl alcove). The graph ${\mathcal A}_4(SU(4))$ is displayed on figure \ref{dimA4}.
   
\subsection{The module and its quantum graph}
   
The next ingredient is an additive category ${\mathcal E}(G)$, not necessarily monoidal, on which the previous one ${\mathcal A}_k(G)$ (which is monoidal) acts, \ie we are given a (monoidal) functor from ${\mathcal A}_k$ to the monoidal category of endofunctors of ${\mathcal E}$. The reader can think of this situation as being an analogue of the action of a group on a given space. ${\mathcal E}$ is called a 
``module category'' \cite{Ostrik,EtingofOstrik}, but the word ``module'' has so many meanings that it may be better to say that we have an action, or that ${\mathcal E}$ is an actegory (a substantive coined by R. Street). It may be sometimes interesting to think that ${\mathcal E}$ can be acted upon in more than one way, so that we can think of the action of ${\mathcal A}_k$ as a particular ``enrichment'' of ${\mathcal E}$.  Irreducible objects of ${\mathcal E}$  are boundary conditions for the corresponding Conformal Field Theory specified by ${\mathcal A}_k$.  It is useful to assume, from now on, that the category ${\mathcal E}$ is indecomposable (it is not equivalent to the direct sum of two non trivial categories with  ${\mathcal A}_k$ action). Like in the classical situation, we have a restriction functor ${\mathcal A}_k \hookrightarrow {\mathcal E}$ and an induction functor ${\mathcal A}_k \hookleftarrow {\mathcal E}$.
  
Since ${\mathcal E}$ is additive, we have a Grothendieck  group, also denoted by the same symbol. Because of the existence of an action, this (abelian) group is a module over the Grothendieck ring of ${\mathcal A}_k$, and it is automatically a $\ZZ_+$ module: the structure constants of the module, usually called annulus coefficients in string theory articles, or in \cite{FuchsRunkelSchweigert-I}, and described by (annular) matrices $F_n = (F_n)_{ab}$, are non negative integers. The index $n$ is a Young  diagram describing an irreducible object (vertices) of ${\mathcal A}_k(G)$, and $a,b$ describe simple objects (vertices) of  ${\mathcal E}$. To the fundamental representations of $G$ correspond particular annular matrices that can be considered as adjacency matrices of a graph (actually we obtain several graphs with the same vertices but various types of edges), that we call the McKay graph of the category ${\mathcal E}$, or simply ``the quantum graph'', for short.  We may think of the Grothendieck group of  ${\mathcal E}$ as the vector space spanned by the vertices of the graph. 
 It is often convenient to introduce a family of rectangular matrices called ``essential matrices'' \cite{Coque:Qtetra} , via the relation $(E_a)_{nb} = (F_n)_{ab}$, and  when $a$ is the origin\footnote{A particular vertex of ${\mathcal E}$ is always distinguished } $0$ of the graph, $E_0$ is usually called ``the intertwiner''.
 
Quantum graphs of type $G=SU(2)$ are the  (simply laced) $ADE$ Dynkin diagram, those of type $G=SU(3)$ were introduced by 
\cite{DiFrancescoZuber}. Existence of the corresponding categories was shown by Ocneanu \cite{Ocneanu:Bariloche}. Classification of $SU(4)$ module categories is also claimed to be completed \cite{Ocneanu:Bariloche}.

The rigidity property of  ${\mathcal A}_k$ implies that the module ${\mathcal E}$ is rigid (or based \cite{Ostrik}). 
In other words:  $ (F_{\overline n})_{ab} = (F_ n)_{ba}$.  This property 
excludes the non-simply laced cases of $G=SU(2)$ type, since  ${\overline \lambda_n} = \lambda_n$. The same property holds but does not exclude double lines for $G=SU(3)$ or higher, so that it is not appropriate to say that higher analogues of $ADE$ Coxeter-Dynkin diagrams are ``simply laced''.

Let us  mention that simple objects $a,b, \ldots$ of the module category  ${\mathcal E}$ can also be thought  as right modules over a Frobenius  algebra  ${\mathcal F}$, which is a particular object in the monoidal category ${\mathcal A}_k$, and which plays an important role in other approaches \cite{Ostrik,FuchsRunkelSchweigert-I}, but we shall not describe its structure here. 

 As already mentioned, the category  ${\mathcal E}$ is not required to be monoidal, but there are cases where it is, so that it has a tensor product, compatible with the ${\mathcal A}_k$ action. In another terminology, one says that the corresponding graphs have self - fusion or that they define ``quantum subgroups'' of $G$, whereas the others are only ``quantum modules''. When it exists, the self-fusion is described, at the level of the module, by another family of matrices $G_a$ with non negative integer entries: we write $a \cdot  b =  \sum_c (G_a)_{bc}\, c$ and compatibility with the fusion algebra reads $\lambda_n \cdot (a \cdot b) = (\lambda_n \cdot a) \cdot b$, so that $(G_a \cdot F_n) = \sum_c (F_n)_{ac} \, G_c $.

\subsection{Quantum symmetries}

The third and final ingredient is the centralizer category of ${\mathcal E}$ with respect to the action of ${\mathcal A}_k$. It is defined as the category of module functors $\mathfrak{f}$ from ${\mathcal E}$ to itself, commuting with the action of ${\mathcal A}_k$, \ie such that $\mathfrak{f}(\lambda_n \otimes \lambda_a)$ is isomorphic with $\lambda_n \otimes \mathfrak{f}(\lambda_a)$, for $\lambda_n \in Ob({\mathcal A}_k)$ and $\lambda_a \in Ob({\mathcal E})$,  via a family of  morphisms $c_{\lambda_n, \lambda_m}$ obeying triangular and pentagonal constraints. We simply call  ${\mathcal O} = Fun_{{\mathcal A}_k}({\mathcal E},{\mathcal E})$ this centralizer category\footnote{For $SU(2)$, the structure of $Fun_{{\mathcal A}_k}({\mathcal E}_1,{\mathcal E}_2)$, where 
${\mathcal E}_{1,2}$ can be distinct module categories was obtained by \cite{Ocneanu:Unpublished}.
}, but one should remember that its definition involves both  ${\mathcal A}_k$ and  ${\mathcal E}$. 

${\mathcal E}$ is  both a module category over ${\mathcal A}_k$ and over ${\mathcal O}$. The later is additive, semisimple and monoidal. The Grothendieck group of ${\mathcal E}$ is therefore not only a  $\ZZ_+$   module over the fusion ring, but also a $\ZZ_+$   module over the Grothendieck ring of ${\mathcal O}$,  called the Ocneanu ring (or algebra) of quantum symmetries and denoted by the same symbol. Structure constants of the ring of quantum symmetries are encoded by matrices $O_x$, called ``matrices of quantum symmetries'';  structure constants of the module, with respect to the action of quantum symmetries,  are encoded by the so called ``dual annular matrices'' $S_x$.

 To each fundamental irreducible representation of $G$ one associates two fundamental generators of ${\mathcal O}$, called chiral (left or right).  For instance, for $G=SU(4)$ quantum graphs, ${\mathcal O}$ has $6 = 2 \times 3$ chiral generators. Like in usual representation theory, all  other linear generators of this algebra  appear when we decompose  products of fundamental (chiral) generators.
  The Cayley graph of multiplication by the  chiral generators (several types of lines),  called the Ocneanu graph of ${\mathcal E}$, encodes the algebra structure of ${\mathcal O}$. 

Quantum symmetries that appear in the decomposition of products of left (right) generators span a subalgebra called the left (right) chiral subalgebra. The chiral subalgebras are not necessarily commutative but the left and the right commute. Intersection of left and right chiral subalgebras is called ambichiral subalgebra. In the particular case ${\mathcal E} = {\mathcal A}_k$, left and right can be identified.  Determining all quantum symmetries can be an arduous task, even in relatively simple situations. A simpler problem is to determine the chiral generators and the Ocneanu graph. We shall give an example later, in the case of a particular exceptional quantum graph of $SU(4)$ type.

 \subsection{The quantum groupo\"\i d}
 
To  ${\mathcal E}$ one can associate a finite dimensional weak bialgebra (or quantum groupo\"\i d) ${\mathcal B}$, which is such that the category ${\mathcal A}_k$ can be realized as $Rep({\mathcal B})$, and also such that the  category ${\mathcal O}$ can be realized as $Rep({\widehat{\mathcal B}})$ where $\widehat{\mathcal B}$ is the dual of ${\mathcal B}$. These two algebras are finite dimensional (actually semisimple in our case) and one algebra structure (say $\widehat{\mathcal B}$) can be traded against a coalgebra structure on its dual. ${\mathcal B}$ is a weak bialgebra, not a bialgebra, because $\Delta \one \neq \one \otimes \one$, where $\Delta $ is the coproduct in  ${\mathcal B}$, and $\one$ is its unit. In our case, it is not only a weak bialgebra but a weak Hopf algebra (we can define an antipode, with the expected properties \cite{BohmSzlachanyi,NikshychVainerman,NikshychTuraevVainerman,Nill}).  

Introducing  a star operation and a scalar product allow identification of ${\mathcal B}$ and its dual,  so that both products (say ``product'' and ``convolution product'') can be defined on the same underlying vector space.
One construction of this bialgebra was given in \cite{Ocneanu:paths}, using a formalism of operator algebras. A corresponding categorical construction is as follows:
Label irreducible objects of categories ${\mathcal A}_k$ by $\lambda_n,\lambda_m,\dots$, of ${\mathcal E}$ by $\lambda_a,\lambda_b,\dots$ and of ${\mathcal O}$ by $\lambda_x, \lambda_y,\dots$. Call $H_{ab}^n = Hom(\lambda_n\otimes \lambda_a, \lambda_b)$, the ``horizontal space of type $n$ from $a$ to $b$'' (also called space of essential paths of type $n$ from $a$ to $b$, space of admissible triples, or triangles). Call $V_{ab}^x = Hom(\lambda_a\otimes \lambda_x, \lambda_b)$ the ``vertical space of type $x$ from $a$ to $b$''. We just take these horizontal and vertical spaces as vector spaces and consider the graded sums $H^n = \sum_{ab} H_{ab}^n$ and $V^x = \sum_{ab} V_{ab}^x$. To construct the weak bialgebra, we take the (graded)  endomorphism algebras ${\mathcal B} = \sum_n End(H^n)$ and  $\widehat{{\mathcal B}} = \sum_x End(V^x)$. For obvious reasons, ${\mathcal B}$ and $\widehat{{\mathcal B}}$ are sometimes called ``algebra of double triangles''. Existence of the bialgebra structure (compatibility) rests on the properties of the pairing, or equivalently on the properties of the coefficients\footnote{Constructions of ${\mathcal B}$, inspired from \cite{Ocneanu:paths}, and using these properties, were given in \cite{PetkovaZuber:Oc} and \cite{CoqueTrinchero:cells}.} (Ocneanu cells) obtained by pairing two bases of matrix units\footnote{Definition of cells involve normalization choices:  the spaces $H_{ab}^n$ are not always one-dimensional, moreover  one may decide to use bases made of vectors proportional to matrix units rather than  matrix units themselves.} for the two products. 
Being obtained by pairing double triangles, Ocneanu cells (generalized quantum $6J$ symbols)  are naturally associated with tetrahedra with two types (black ``b'', or white ``w'') of vertices, so that edges $bb$, $bw$ or $ww$ refer to labels $n$, $a$, $x$ of ${\mathcal A}$, ${\mathcal E}$ and ${\mathcal O}$. 

The ${\mathcal A}_k \times {\mathcal O}$ module category ${\mathcal E}$ can be recovered from the study of the source and target subalgebras of ${\mathcal B}$, but in practice it is often obtained by first determining the graph of quantum symmetries from the solution of the so-called ``modular splitting equation'' (see later), which involves only a single piece of data: the modular invariant.

\subsection{Torus structure and modular splitting equation}

From results obtained in operator algebra by \cite{Ocneanu:Unpublished} and  \cite{Evans-I, Evans-II, Evans-Kawahigashi}, translated to a categorical language by 
\cite{Ostrik}, one can show that 
existence of a braiding in the category ${\mathcal A}_k$ leads to the bimodule property  $ {\mathcal A}_k \times {\mathcal O} \times {\mathcal A}_k  \mapsto  {\mathcal O}$, and this reads, at the level of Grothendieck rings, $m \, x \, n = \sum_y \, (W_{x,y})_{mn} \, y$, where $m,n$ refer to irreducible objects of  ${\mathcal A}_k$, $x,y$ to irreducible objects of ${\mathcal O}$, and where $W_{x,y}$ constitute a family of so - called toric matrices, with matrix elements $(W_{x,y})_{mn}$, again non negative integers.  

When both $x$ and $y$ refer to the unit object (that we label $0$), one recovers the modular invariant ${\mathcal M}=W_{0,0}$ encoded by the partition function $\mathcal{Z}$ of conformal field theory.  As explained in \cite{PetkovaZuber:Oc}, when one or two indices $x$ and $y$ are non trivial, toric matrices are interpreted as partition functions on a torus,  in a conformal theory of type $ {\mathcal A}_k$,  with boundary type conditions specified by ${\mathcal E}$,  but with defects specified by $x$ and $y$. Only ${\mathcal M}$ is modular invariant (it commutes with the generator $s$ and $t$ of $SL(2,\ZZ)$ in the Hurwitz - Verlinde representation). Toric matrices were first introduced and calculated by Ocneanu (unpublished) for theories of type $SU(2)$. Various methods to compute or define them can be found in \cite{FuchsRunkelSchweigert-I,Coque:Qtetra,PetkovaZuber:Oc}. Reference  \cite{GilCoque:ADE} gives explicit expressions for all $W_{x,0}$, for all members of the $SU(2)$ family ($ADE$ graphs).

Left and right associativity constraints $(m(n x p) q)=(mn)x(pq)  $ for the $ {\mathcal A} \times {\mathcal A}$ bimodule structure of ${\mathcal O}$ can be written in terms of fusion and toric matrices; a particular case of this equation reads
$
\sum_x (W_{0,x})_{\lambda\mu} \, W_{x,0} = N_{\lambda}\, {\mathcal M}  \,N_{\mu}^{tr} \;.
$
It was presented by A.Ocneanu in \cite{Ocneanu:Bariloche} and called the  ``modular splitting equation''.
A practical method to solve it is given in \cite{EstebanGil}, with several $SU(3)$ examples.
 Given fusion matrices $N_p$ (known in general) and a modular invariant matrix ${\mathcal M}=W_{0,0}$, 
solving this equation, \ie finding the $W_{x,0}$, allows one, in most cases, to construct the chiral generators of  ${\mathcal O}$ and obtain the graph of quantum symmetries. 

\subsection{Triangular cells,  self-connection on ${\mathcal E}$ and pocket equations}

Given a graph defining a module over a fusion ring ${\mathcal A}_k$ for some Lie group $G$,  the question is to know if it is a ``good graph'', ie, if the corresponding module category indeed exists. According to A. Ocneanu \cite{Ocneanu:Bariloche}, when the rank is $\geq 2$, this will be the case if and only if  one can associate, in a coherent manner,  a complex number to each triangle of the graph : this defines (up to some kind of gauge choice)   {\sl a self-connection on the set of triangular cells}. 
There are two compatibility  equations  respectively called the small and the large  {\sl pocket equations}, for these triangular cells.

\subsection{Summary}

Obtaining the list of all quantum graphs of type $G$ (all module categories of type $G$), together with their various properties, often requires a number of complementary techniques, and it may not be possible to propose a multi-purpose machinery that would work in all cases.  It is clearly always possible to  define the so-called ``diagonal cases'': $ {\mathcal E} =  {\mathcal A}_k (G)$. Then, 
using orbifold techniques, one can build infinite series $ {\mathcal E} = {\mathcal D} $ generalizing the even or odd $D$ diagrams of $SU(2)$ (which are $\ZZ_2$ orbifolds of the $A$ diagrams at the same level). The difficulty is to obtain the exceptional cases. One argument, due to A. Ocneanu (unpublished), shows that the number of exceptional cases corresponding to a given Lie group $G$ is finite. The strategy followed to determine and study an exceptional case ${\mathcal E}$ can be summarized as follows.

\begin{enumerate}
\item[$\bullet$] Choose a Lie group $G$ and a level $k$. Determine the fusion matrices $N_p$ from the adjacency matrix of ${\mathcal A}_k (G)$ and representation theory of $G$ truncated at level $k$ (known recurrence formulae).
\item[$\bullet$]  Determine the generators $s$ and $t$ for the Hurwitz-Verlinde representation of $SL(2,Z)$.
\item[$\bullet$] Choose a modular invariant. It can be obtained, either from arithmetic considerations, or from the existence of a conformal embedding.
\item[$\bullet$] Solve the modular splitting equation, \ie determine the toric matrices $W_{x,0}$.
\item[$\bullet$] Determine the chiral generators of the algebra of quantum symmetries and the Ocneanu graph.
\item[$\bullet$] Find possible candidates for the quantum graph of ${\mathcal E}$: it is usually (but not always)  a component of the graph of quantum symmetries. Check that a self-connection, for this chosen solution, indeed exists. 
\item[$\bullet$] Now that ${\mathcal E}$ is known, one can determine the annular matrices $F_n$ that encode the module structure over the fusion algebra ${\mathcal A}_k$. In turn, these matrices encode several other quantities of interest, in particular the induction-restriction rules, the (quantum) dimensions, and the size of the simple blocks of the quantum groupo\"\i d ${\mathcal B}$ for its first multiplicative structure.
\item[$\bullet$] One can investigate whereas the quantum graph under study possesses self-fusion and determine, in that case,  the so-called graph matrices $G_a$.
\item[$\bullet$] Determination of the full structure of  the quantum groupo\"\i d ${\mathcal B}$, in particular the pairing between  ${\mathcal B}$ and $\widehat{{\mathcal B}}$,  is out of reach in general since it would involve a calculation of all Ocneanu cells, and this seems to be intractable. However in many cases, it is possible to exhibit a set of linear generators $O_x$ for the algebra of quantum symmetries ${\mathcal O}$ (remember that chiral generators have been obtained in a previous step). From this, one can determine the dual annular matrices $S_x$ encoding the module structure of ${\mathcal E}$ over ${\mathcal O}$ without having to rely on an explicit determination of the pairing.
 As a by-product, one obtains the size of the simple blocks of the dual of  ${\mathcal B}$ for its multiplicative structure.
\end{enumerate}

\subsection{ Classification results }

In the case of $SU(2)$, we have the the $ADE$ classification of \cite{CIZ}.  Self-fusion exists only in the cases of graphs $A_r$ (then $k=r-1$),  $D_{even}$ (then $k=0 \,  \mbox {\rm mod\/} \, 4$), $E_6$, then $k=10$, and $E_8$, with $k=28$. The cases $D_{odd}$, with levels $k=2 \, \mbox {\rm mod \/} 4$,  and $E_7$, at level $16$ do not admit self-fusion.
In the case of $SU(3)$ we have the Di Francesco - Zuber diagrams \cite{DiFrancescoZuber} and 
the  classification of modular invariants of type $SU(3)$ by \cite{Gannon-SU3}. Notice that, sometimes, the same modular invariant can be associated with distinct module categories (distinct quantum graphs).  Existence of the corresponding categories was shown by A. Ocneanu \cite{Ocneanu:Bariloche}, actually one of the candidates (``fake graph'') had to be discarded because it did not pass the self-connection test. Several $SU(3)$ graphs have self-fusion, namely:  ${\mathcal A}_k$ itself, the ${\mathcal D}_k$ (whose McKay graphs are $\ZZ_3$ orbifolds of those of ${\mathcal A}_k$) when $k$ is divisible by $3$, and three exceptional cases called ${\mathcal E}_5$, ${\mathcal E}_9$ and ${\mathcal E}_{21}$, at levels $5$, $9$ and $21$. 
The other quantum graphs (no self-fusion) are: the series ${\mathcal A}_k^c$, for which the number of simple objects is equal to the number of self dual simple objects in  ${\mathcal A}_k$, the ${\mathcal D}_k$ series, when $k=1$ or $2$ mod $3$, the series ${\mathcal D}_k^c$, for all $k$, two modules of exceptionals called ${\mathcal E}_5/3$,
${\mathcal E}_9/3$, and finally the exceptional case ${\mathcal D}_9^t$ (a generalization of $E_7$ that can be obtained from ${\mathcal D}_9$ by an exceptional twist), along with its ``conjugate case'' called  ${{\mathcal D}_9}^{tc}$. Useful information can be found in \cite{GilDahmaneHassan}.

A classification of $SU(4)$ quantum graphs was presented by A. Ocneanu in Bariloche  \cite{Ocneanu:Bariloche}.  One finds the ${\mathcal A}_k$ series, with two kinds of orbifolds (the ${\mathcal D}^{(2)}_{k} = \mathcal{A}_k/2$ and ${\mathcal D}^{(4)}_{k} = \mathcal{A}_k/4$ series, with
self-fusion for $\mathcal{A}_k/2$ when $k$ is even and for $\mathcal{A}_k/4$ when $k$ is divisible by 4), together with their conjugates, three exceptional quantum graphs with self-fusion, at levels $4$, $6$ and $8$, denoted ${\mathcal E}_4, {\mathcal E}_6, {\mathcal E}_8$ together with one exceptional module for each of the last two, and finally one exceptional ${\mathcal{D}^{(4)t}_{8}}$ case without self-fusion (again a generalization of $E_7$), along with a conjugate graph called ${\mathcal{D}^{(4)tc}_{8}}$. The exceptional modular invariants at level $4$ and $6$ were found by \cite{SchellekensYankielowicz, AltschulerBauerItzykson}, and at level $8$ by \cite{AldazabalEtAl}. The corresponding quantum graphs ${\mathcal E}_4$,  ${\mathcal E}_6$ and ${\mathcal E}_8$ were respectively determined by \cite{PetkovaZuberNP1995, PetkovaZuberNP1997, Ocneanu:Bariloche}.

For higher rank, infinite series of examples can be obtained from the ${\mathcal A}_k$ graphs, using  conjugacies, orbifold techniques (generalizations of the $D$ graphs) or semi-simple (non simple) conformal embedding followed by contraction.  Many exceptional examples can also be obtained, sometimes thanks to the existence of conformal embeddings, or using rank-level duality considerations. One can also follow arithmetical approaches (study  the commutant of $SL(2,\ZZ)$), this leads to modular invariants \cite{Gannon-SU3,Gannon-SU3-bis} that  become candidates as possible partition functions, but their number often exceeds, by far, the ``physical'' ones, which are those associated with quantum graphs of type $G$.


\section{Conformal embeddings}
 
\subsection{Generalities}
 
For our purposes, it will be enough to consider the following situation : 
Take $K$,  a simple Lie group of dimension $d_K$ and  $G$ a Lie subgroup of dimension $d_G$.
Denote by ${\mathfrak k}$ and ${\mathfrak g} $ their Lie algebras. 
Call  $S = K/G$ the corresponding homogenous space. Write ${\mathfrak k}  = {\mathfrak g} \oplus {\mathfrak s}$.
 The Lie group $G$ acts on the vector space ${\mathfrak s}$ by the isotropy representation, which is usually reducible, and one obtains its decomposition into irreducible components ${\mathfrak s} = \oplus_i  {\mathfrak s}_i$ by reducing the adjoint of $K$ with respect to the adjoint of $G$. The group
 $K$ being simple, it has a unique Killing form, up to scale, and there is a canonical normalisation for all simple Lie groups. One can use this form to define, for each component ${\mathfrak s}_i$ a Dynkin index $k_i$. An embedding is therefore associated with a sequence of numbers $k_1, k_2 \ldots$ Assuming now that $G$ is a maximal subgroup of $K$  such that $K/G$ is irreducible  (more precisely ``isotropy irreducible'' \ie such that the representation ${\mathfrak s}$ of the group $G$ is real irreducible\footnote{Notice that  $SU(3) \subset SU(5)$,  for instance, is not irreducible ($SU(5)$ does not possess any $SU(3)$ subgroup that could be maximal) whereas there is an embedding $SU(3) \subset SU(6)$ which is maximal and irreducible since  $35 \hookrightarrow 8 + 27$ (the representation $27$ being indeed real and irreducible). The representation  ${\mathfrak s}$ is real irreducible, although it may be reducible on the field of complex numbers. For instance, if we consider the embedding of $SU(3)$ in $SO(8)$ obtained from the adjoint representation, we have $28 \hookrightarrow 8 \oplus 10 \oplus \overline{10}$ and the isotropy representation is not complex irreducible (it is $10 \oplus \overline{10}$), but it is real irreducible, of dimension $20$, and the real manifold $SO(8)/SU(3)$ is indeed irreducible.
}), we have then a unique value for the Dynkin index $k$ of the embedding.
The embedding is conformal if $k$ is an integer and if the following identity is satisfied :
\begin{equation}
 \frac{d_G \,  k }{ k + g_G} = \frac{d_K }{ 1 + g_K}
 \label{centralcharge}
\end{equation}
 where $g_K$ and $g_G$ are the dual Coxeter numbers of $K$ and $G$. 
 One denotes by  $c$ the common value of these two expressions.
When $K$ is simply laced, $c$ is equal to its rank.
This  definition does not require the framework of affine Lie algebras to make sense, but it is justified from the fact that, given an embedding ${\mathfrak g}   \subset {\mathfrak k} $ of Lie algebras, and an irreducible unitary highest weight module  of the affine algebra $\widehat {\mathfrak k}$ at some level,  one obtains a set of representations of the Virasoro algebra which intertwines with the action of $\widehat  {\mathfrak g}$ on the same module (in other words, we are computing multiplicities) and the embedding is called conformal when those multiplicities are finite, something which occurs precisely when the level of $\widehat {{\mathfrak k}}$ is $1$ and when the above identity between central charges, at respective levels $k$ (for $\widehat{\mathfrak g}$), and $1$ (for $\widehat{\mathfrak g}$),  is satisfied. It would be interesting to interpret the same condition in terms of (small) quantum groups at roots of unity. In this respect,  notice that level $k$, for $SU(N)$, reads  $q^{k+ N}=1$,  in terms of roots of unity.
The study of embeddings of affine Lie algebras is a rather old subject and we shall only mention a few  ``historical'' references :  \cite{SchellekensWarner,BaisBouwknegt,KacWakimoto,AltschulerBauerItzykson,PDO}, see also \cite{YellowBook}.  Conformal embeddings are also used in the context of subfactors, see for instance \cite{Wassermann,FengXu}.

Warning: One can sometimes find the claim that the previous identity between central charges (at respective levels $k$ and $1$) provides a necessary and sufficient condition for the existence of a conformal embedding, $k$ being then interpreted as the Dynkin index of the corresponding finite dimensional Lie algebras. This should however be taken with a grain of salt, since one should check that there exists indeed a homogeneous space $K/G$ whose Dynkin index, for the embedding $G \subset K$, is equal to the given value $k$. This condition will be called the ``irreducibility requirement''.
 Actually, if we choose a priori some simple Lie group $G$, it is rather easy to solve equation  (\ref{centralcharge}) over the positive integers, \ie we can find a finite family of solutions with $K$ a simple Lie group, and $k$ a positive integer. Take for instance $G=SU(3)$,  one finds immediately that $K$ could be equal to
{\small $SU(5), SU(6), SU(7),SU(8), Spin(8),Spin(10), Spin(12), Spin(14)$},  {\small $ E_6, E_7, Spin(13), Spin(15), Sp(4), Sp(6)$} for appropriate values of the level $k$ of $G$. But if we impose moreover the irreducibility requirement (so that $K/G$, with given Dynkin index $k$, indeed exists), only the following solutions are left: $K=SU(6), Spin(8), E_6, E_7$, at levels $k=5,3,9,21$ for $G$. Only those solutions define conformal embeddings and their associated modular invariant gives rise to the quantum graphs ${\cal E}_5,{\cal D}_3,{\cal E}_9$ and ${\cal E}_{21}$ of $SU(3)$ type.

\subsection{Classification issues}

In order to find exceptional quantum groupo\"\i ds of type $G$, we could give ourselves a Lie group $G$ and try to  embed it conformally in a larger group $K$. One can certainly use results from the (old) literature (that often proceeds from $K$ to $G$ rather than the converse), but for -- given -- simple\footnote{Simplicity of $G$ is  a strong hypothesis that, of course, does not cover all interesting conformal embeddings.}  $G$ it is easy to solve equation (\ref{centralcharge}) using standard results on Lie groups.

Case $SU(2)$. Equation (\ref{centralcharge}) admits 3 solutions and the irreducibility requirement does not change this result. At levels $k= 4, 10, 28$ one finds $K=SU(3), Spin(5)$ and $G_2$. Their associated modular invariant gives rise to the quantum graphs $D_6= {\cal D}_4 (SU(2))$, $E_6={\cal E}_{10} (SU(2))$ and $E_8=  {\cal E}_{28} (SU(2))$.

Case $SU(3)$.  Equation (\ref{centralcharge})  admits 14 solutions (see above) but the irreducibility requirement brings down this number to 4, at levels $k=3,5,9,21$,  namely $K= Spin(8),SU(6), E_6$ and $ E_7$. Their associated modular invariant gives rise to the quantum graphs ${\cal D}_3(SU(3))$, ${\cal E}_5(SU(3))$, ${\cal E}_9(SU(3))$ and ${\cal E}_{21}(SU(3))$.

Case $SU(4)$. Equation (\ref{centralcharge})  admits $21$ solutions  but the irreducibility requirement brings down this number to 4, at levels $k= 2, 4, 6,8 $, namely $K=SU(6), Spin(15)$, $SU(10)$ and $Spin(20)$. Their associated modular invariant gives rise to the quantum graphs $\mathcal{D}_2= \mathcal{A}_2/2(SU(4)),  \, {\mathcal E}_4(SU(4))$, ${\mathcal E}_6(SU(4))$ and ${\mathcal E}_8(SU(4))$.

The following regular series of inclusions are always conformal
\begin{subeqnarray}
SU(N)_{N-2} &\subset& SU(N(N-1)/2) \\
SU(N)_{N+2} &\subset& SU(N(N+1)/2) \\
SU(N)_N &\subset& Spin(N^2-1)
\end{subeqnarray} 
The last inclusion being actually a particular case of the conformal embedding $G_{g_G} \subset Spin(d_G)$. 
For $N=2$, we have only the second one $SU(2)_4 \subset SU(3)$, the other two embeddings of $SU(2)$ 
being ``truly exceptional'' in the sense that they do not belong to these regular series of inclusions. 
For $N=3$ only the second and third case of these regular series of inclusions do exist, namely $SU(3)_5 \subset SU(6)$ and $SU(3)_3 \subset Spin(8)$; the two others embeddings of $SU(3)$ are truly exceptional.
For $N=4$ and above, the three members of the regular series exist. Note that in the case of $SU(4)$ one does not even find a ``truly exceptional'' conformal embedding, in this sense.

It was written in \cite{Ocneanu:Bariloche} that  ${\cal E}_8(SU(4))$ does not seem to correspond to any conformal embedding but we found that it could actually be obtained from $SU(4) \subset SO(20)$ and later discovered (!) that this had been known long ago \cite{AldazabalEtAl}.  We shall return in the conclusion to the modular invariants of the $SU(4)$ family.
For the $SU(5)$ family, we shall have regular exceptional graphs at levels $3$, $7$ and $5$. Notice that  ${\cal E}_3 (SU(5))$ is dual\footnote{Rank-level duality property: if we have a conformal embedding at level $k$,  $SU(N) \subset K$ , there is also a conformal embedding of $SU(k)$, at level $N$, in some appropriate Lie group.}  of ${\cal E}_5(SU(3))$ and that  ${\cal E}_5(SU(5))$ is self-dual.

Remark:
The graph $E_7 =  {\cal E}_{16} (SU(2))$ is not associated with a conformal embedding of the type  $SU(2) \subset K$, and it does not enjoy self-fusion,  but there exists a conformal embedding at level $16$ of $SU(2)_{16} \times SU(3)_6 \subset (E_8)_1$:   $c(SU(2)_{16})=8/3$ and $c(SU(3)_{6})=16/3$ so that the sum of both is indeed  $c(E_8)_{1} = 8$. Therefore one would expect to find an exceptional quantum graph associated with the non simple Lie group  $SU(2)\times SU(3)$, the graph $E_7$ being obtained\cite{YellowBook} from the later by contraction (followed by a subtraction involving the quantum graph $D_{10}$). Of course, there exist higher analogues of this phenomenon.

\subsection{Modular invariants}

\paragraph{Reminder: Modular matrices $s$ and $t$.}
Call $\mathfrak g$ a Lie algebra of rank $r$. Call $\alpha_i$ the simple roots of $\mathfrak g$ ($i=1,\dots,r$), $\alpha_i^{\vee}$ the coroots and $w_i$ the fundamental weights, which obey $\langle w_i,\alpha_i^{\vee}\rangle=\delta_{ij}$. A weight $\lambda$ is written $\lambda = \sum_{i} \lambda_i w_i = (\lambda_1,\lambda_2,\dots,\lambda_r)$. Call $\rho$ the Weyl vector $\rho=\sum_i w_i = (1,1,\dots,1)$. Call $W$ the Weyl group. 
Matrix expressions for representatives of the generators $s$ and $t$ of a double cover of $PSL(2,\mathbb{Z})$, at level $k \in \mathbb{Z}_+$, are given by \cite{Kac-Peter}:
\begin{subeqnarray}
(t)_{\lambda\mu} &=& \exp \left(2 i \pi \left( \frac{|\lambda+\rho|^2}{2\kappa} - \frac{|\rho|^2}{2 g} \right) \right) \delta_{\lambda\mu} = \exp \left( 2 i \pi m_{\lambda}\right) \delta_{\lambda\mu}\\
(s)_{\lambda\mu} &=& \sigma \, \kappa^{-r/2} \sum_{w \in W} \epsilon(w) \exp \left(\frac{2 i \pi \langle w(\lambda+\rho),\mu+\rho\rangle}{\kappa} \right)
\end{subeqnarray} 
where $g$ is the dual Coxeter number, the altitude $\kappa = k + g $, $\epsilon(w)$ is the signature of the Weyl permutation and $\sigma$ a coefficient defined by 
$\sigma = i^{|\Delta_+|} \, (det(\alpha_i^{\vee},\alpha_j^{\vee}))^{-1/2}$, where $|\Delta_+|$ is the number of positive simple roots. $s$ and $t$ matrices are unitary and satisfy 
$(s\,t)^3 = s^2 = \mathcal{C}$, the charge conjugation matrix satisfying $\mathcal{C}^2=\munite$.  
For $SU(4)$, we have $g=4$, $|\Delta_+|=3$, the simple Weyl reflections are 
{\small \begin{eqnarray*}
s_1(\lambda_1,\lambda_2,\lambda_3) &=& (-\lambda_1,\lambda_1+\lambda_2,\lambda_3) \\
s_2(\lambda_1,\lambda_2,\lambda_3) &=& (\lambda_1+\lambda_2,-\lambda_2,\lambda_2+\lambda_3) \\
s_3(\lambda_1,\lambda_2,\lambda_3) &=& (\lambda_1,\lambda_2+\lambda_3,-\lambda_3)
\end{eqnarray*}}
and the full Weyl group, of order 24, is generated by products of $s_i$ with $s_i^2=1$,  $(s_1s_2)^3=1$, $(s_1s_3)^2=1$, $(s_2s_3)^3=1$. 
The scalar product of two weights is:
{\small $$
\langle \lambda , \mu \rangle = \frac{1}{4}\left( \lambda_1 (3\mu_1 + 2\mu_2 + \mu_3) + 2\lambda_2 (\mu_1 + 2\mu_2 + \mu_3) + \lambda_3 (\mu_1 + 2\mu_2 + 3\mu_3) \right) 
$$}
Using this, one finds explicit expressions for $s$ and $t$ matrices. 
The $t$ matrix obeys $t^{8 \kappa} = \munite$.

\paragraph{The method.} Obtaining  modular invariants  from  conformal embeddings $G_k \subset \widehat K_1$ is explained for instance in \cite{YellowBook} and we only summarize part of this information here. 

Using the language of affine Lie algebra, one has first to determine the integrable irreducible highest weight representations (i-irreps for short, from now on) $\lambda$ at the chosen level. First of all, the level should be big enough\footnote{in terms of quantum groups at roots of unity this means that if the root of unity is too small, there will be no irreducible representation of non vanishing $q$-dimension.}: this  integrability condition reads 
\begin{equation}
k \geq \langle \lambda, \theta \rangle
\label{intcondition}
\end{equation} 
where $\theta$ is the highest root of the chosen Lie algebra.
To such an i-irrep $\lambda$ one associates a conformal dimension defined by 
\begin{equation}
h_{  \lambda} = \frac{ \langle \lambda, \lambda + 2 \rho \rangle}{2(k+g)}
\label{confdim}
\end{equation} 
where $k$ is the level,  $g$ is the dual Coxeter number of the chosen  Lie algebra, $\rho$ is the Weyl vector and $h_{  \lambda}$ is related to the phase $m_{\lambda}$ of the $t$ matrix by $m_{\lambda} = h_{\lambda} - c/24$. 
We make the list of i-irreps $\lambda$  of $K$ at level $1$ and calculate their conformal dimensions $h_{\lambda}$.
We make the list of i-irreps $\mu$ of $G$ at level $k$ and calculate their conformal dimensions $h_{\mu}$.
A necessary -- but not sufficient -- condition for an (affine or quantum) branching from $\lambda$ to $\mu$ is that  $h_{\lambda} = h_{\mu} + n$ 
for some positive integer $n$.
So we can make a list of candidates ${  \lambda} \hookrightarrow \sum_j c_j \, {  \mu}_j $ where $c_j$ are positive integers to be determined.
We write the diagonal invariant of type $K$ as a sum $\sum_s  \lambda_{\overline s} \lambda_s $. It should give rise to a quantum graph of type ${\mathcal A}_1(K)$.
Using the affine branching rules (also quantum branching rules), we replace, in this expression,  each $\lambda_s$ by the corresponding sum of i-irreps for $G$. The modular invariant $\mathcal{M}$ of type $G$ that we are looking for is parametrized by
\begin{equation}
{\mathcal Z} = \sum_s  (\sum_j c_j(\overline s) \, {  \mu}_j (\overline s)) (\sum_j c_j(s) \, {  \mu}_j (s))
\end{equation}
There exist several techniques to determine the coefficients $c_j$, for instance using information coming from the finite branching rules. One method,  that may lack of elegance, but which is quite efficient, is simply to impose that $\mathcal{M}$, parametrized as above,  commutes with the known generators $s$ and $t$ of $PSL(2,\ZZ)$ and to determine the $c_j$ by solving linear equations.

\section{The ${\mathcal E}_4 (SU(4))$ example}

\subsection{Conformal embedding and the modular invariant.}
We are interested in finding and studying an exceptional quantum graph for the $A_3=SU(4)$ system  coming from a conformal embedding. We choose the embedding of  $SU(4)$, at level $k=4$, into $B_7=Spin(15)$, at level $1$.  Using $dim(SU(4))=15$, $g_{SU(4)}=4$, we find  $c = 15/2$. Using $dim(Spin(15))=dim(B_7)= 105$, $g_{Spin(15)}=13$, we check that $c=15/2$ as well. The homogeneous space $Spin(15)/SU(4)$ is isotropy irreducible : reduction of the adjoint representation of $Spin(15)$ with respect to $SU(4)$ reads $105 \mapsto 15 + 90 $, and $90$ is (real) irreducible. Using standard formulae, we check that the Dynkin index of this embedding is equal to $4$.
\begin{itemize}
 
\item
The Cartan matrix of $B_7$  is $2 \,  Id - G[B_7]$, where $G[B_7]$ is the adjacency matrix of the Dynkin diagram of $B_7$. This Lie group is non simply laced, therefore  its quadratic form matrix $Q$ is the inverse of the matrix obtained by multiplying the last line of the Cartan matrix by a coefficient $2$.
\begin{eqnarray*}
G[B_7] =
\left(
\begin{array}{lllllll}
 0 & 1 & 0 & 0 & 0 & 0 & 0 \\
 1 & 0 & 1 & 0 & 0 & 0 & 0 \\
 0 & 1 & 0 & 1 & 0 & 0 & 0 \\
 0 & 0 & 1 & 0 & 1 & 0 & 0 \\
 0 & 0 & 0 & 1 & 0 & 1 & 0 \\
 0 & 0 & 0 & 0 & 1 & 0 & 2 \\
 0 & 0 & 0 & 0 & 0 & 1 & 0
\end{array}
\right)
& \qquad
Q= \left(
\begin{array}{lllllll}
 1 & 1 & 1 & 1 & 1 & 1 & \frac{1}{2} \\
 1 & 2 & 2 & 2 & 2 & 2 & 1 \\
 1 & 2 & 3 & 3 & 3 & 3 & \frac{3}{2} \\
 1 & 2 & 3 & 4 & 4 & 4 & 2 \\
 1 & 2 & 3 & 4 & 5 & 5 & \frac{5}{2} \\
 1 & 2 & 3 & 4 & 5 & 6 & 3 \\
 \frac{1}{2} & 1 & \frac{3}{2} & 2 & \frac{5}{2} & 3 & \frac{7}{4}
\end{array}
\right)
\end{eqnarray*}
An arbitrary weight reads $\lambda = (\lambda_j)$ in the base of fundamental weights, and in this base the Weyl vector is $\rho = (1,1,1,1,1,1,1)$. The scalar product of two weights is  $\langle \lambda, \mu \rangle = (\lambda_i) Q_{ij} (\mu_j)$. 
At level 1, there are only three i-irreps for $B_7$ (use equation (\ref{intcondition})), namely\footnote{We never write explicitly the affine component of a weight since it is equal to $k- \langle \lambda, \theta \rangle$.} $(0)$ , $(1,0,0,0,0,0,0)$ or $(0,0,0,0,0,0,1)$.  
From equation (\ref{confdim}) we calculate their conformal dimensions: $\left\{0,\frac{1}{2},\frac{15}{16}\right\}$.

\item    The Cartan matrix of $A_3=SU(4)$  is $2 \,  Id - G[A_3]$, where $G[A_3]$ is the adjacency matrix of the Dynkin diagram $A_3$.  
Its quadratic form matrix $Q$ is the inverse of the Cartan matrix.
 \begin{eqnarray*}
 G[A_3] =
\left(
\begin{array}{lll}
 0 & 1 & 0 \\
 1 & 0 & 1 \\
 0 & 1 & 0
\end{array}
\right)
& \qquad
 Q=
\left(
\begin{array}{lll}
 \frac{3}{4} & \frac{1}{2} & \frac{1}{4} \\
 \frac{1}{2} & 1 & \frac{1}{2} \\
 \frac{1}{4} & \frac{1}{2} & \frac{3}{4}
\end{array}
\right)
\end{eqnarray*}
In the base of fundamental weights\footnote{We use sometimes the same notation $\lambda_i$ to denote a representation or to denote the Dynkin labels of a weight; this should be clear from the context.}, the Weyl vector is $\rho = (1,1,1)$, and at level $k$ the i-irreps $\lambda = (\lambda_1, \lambda_2, \lambda_3)$ are such that 
$0 \leq \lambda_1 +  \lambda_2 +  \lambda_3 \leq k$, of cardinality $(k+1)(k+2)(k+3)/6$. At level 4, we calculate the $35$ conformal dimensions for $SU(4)$ i-irreps 
using
{\small
$$
h_{\lambda} =\frac{1}{16} \left(({\lambda_1}+2) \left(\frac{3 {\lambda_1}}{4}+\frac{{\lambda_2}}{2}+
\frac{{\lambda_3}}{4}\right)+({\lambda_2}+2) \left(\frac{{\lambda_1}}{2}+{\lambda_2}+
\frac{{\lambda_3}}{2}\right)+({\lambda_3}+2) \left(\frac{{\lambda_1}}{4}+
\frac{{\lambda_2}}{2}+\frac{3 {\lambda_3}}{4}\right) \right)
$$
   }
and find (use obvious ordering with increasing level):
{\tiny
$$ 0,\frac{15}{64},\frac{5}{16},\frac{15}{64},\frac{9}{16},\frac{39}{
   64},\frac{1}{2},\frac{3}{4},\frac{39}{64},\frac{9}{16},\frac{63}{64},1
   ,\frac{55}{64},\frac{71}{64},\frac{15}{16},\frac{55}{64},\frac{21}{16}, \frac{71}{64},1,\frac{63}{64},\frac{3}{2},\frac{95}{64},\frac{21}{16}
   ,\frac{25}{16},
   \frac{87}{64},\frac{5}{4},\frac{111}{64},\frac{3}{2},\frac{87}{64},\frac{21}{16},2,\frac{111}{64},\frac{25}{16},\frac{95}{64}
   ,\frac{3}{2}
$$}

\item  The difference between conformal dimensions of $B_7$ and $A_3$ should be an integer. This selects 
the three following possibilities:
{\small  \begin{eqnarray*}
 0000000 \hookrightarrow^? 000 + 210 + 012 + 040, \quad
 &
 1000000  \hookrightarrow^?  101 + 400 + 121 + 004, \quad
 &
 0000001 \hookrightarrow^? 111.
 \end{eqnarray*}
}
The above three possibilities give only necessary conditions for branching :  all representations on the right hand side do not necessarily appear, or they may appear with multiplicities. One may determine these coefficients, for instance by introducing arbitrary parameters and imposing that the candidate for the modular invariant matrix indeed commutes with the generators $s$ and $t$ of $SL(2,\ZZ)$.
In this way one discovers that the multiplicity of $(111)$ should be $4$, and that all the other coefficients indeed appear, with multiplicity $1$. 
The $35\times 35$ modular invariant matrix $\mathcal{M}_{\lambda\mu}$, or the corresponding partition function\footnote{
Some authors write instead $\mathcal{Z} = \sum \chi_{\lambda}\,\mathcal{M}_{\lambda\overline{\mu}}\,\bar{\chi}_{\mu}$, and therefore some care has to be taken in order to compare results since conjugated cases can be interchanged.} $\mathcal{Z} = \sum_{\lambda}\chi_{\lambda}\,\mathcal{M}_{\lambda\mu}\,\bar{\chi}_{\mu}$ obtained from the diagonal invariant $ \vert 0000000\vert^2 +   \vert 1000000\vert^2 +    \vert 0000001\vert^2$ of $B_7$,  reads :
\begin{equation}
\mathcal{Z}(\mathcal{E}_4) = \vert 000 + 210 + 012 + 040 \vert^2 + \vert 101 + 400 + 121 + 004 \vert^2 + 4 \vert 111 \vert^2
\end{equation}
It introduces a partition on the set of exponents, defined as  the i-irreps corresponding to the non-zero diagonal entries of $\mathcal{M}$ :  $\{ 000, 210,  012, 040, 101, 400, 121, 004, 111\}$.

From the expression of  $\mathcal{Z}$,  we discover that the quantum graph ${\mathcal E}_4 (SU(4))$ has $Tr(\mathcal{M})=12$ vertices but we expect\cite{Ocneanu:Unpublished,Evans-I,Evans-II,Evans-Kawahigashi} $Tr(\mathcal{M}^\dagger \mathcal{M})=48$ quantum symmetries. Because of a coefficient $4$ in $\mathcal{M}$ we expect that the algebra $Oc({\mathcal E}_4)$ spanned by these quantum symmetries is non commutative and possesses a block isomorphic with the algebra of matrices ${ M}(4,\CC)$.
\end{itemize}

\subsection{The quantum graph ${\mathcal E}_4$ and its algebra of quantum symmetries}

\begin{figure}
\centerline{\scalebox{0.5}{\includegraphics{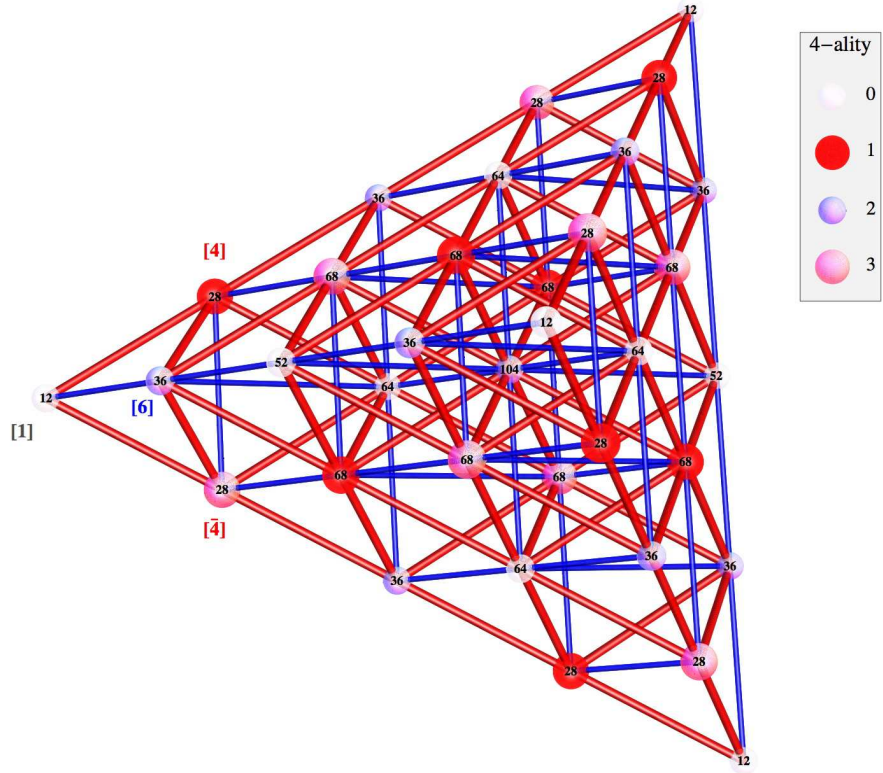}}} 
\caption{The ${\mathcal A}_4$ quantum graph. Generators are marked with their classical dimensions $[4], [6]$ and $[\overline{4}]$. Identity is $[1]$. 
Multiplication by $[4]$ (resp $[\overline{4}]$) is encoded by red edges (thick lines) oriented from vertices of 4-ality $\tau$ to $\tau + 1$ (resp.  $\tau - 1$) modulo $4$. Multiplication by $[6]$ is encoded by unoriented blue edges (thin lines). Numbers in the spheres give the dimension of the simple blocks $n$ of the bialgebra ${\mathcal B}({\mathcal E}_4)$ for its first multiplicative structure.}
\label{dimA4}
\end{figure}

\paragraph{Fusion matrices.}
The trivial representation is $\lambda  = (0,0,0)$ and the three fundamental i-irreps are $(1,0,0), (0,1,0)$ and $(0,0,1)$.
Note that $\lambda = (\lambda_1,\lambda_2,\lambda_3)$ and $\overline{\lambda} = (\lambda_3,\lambda_2,\lambda_1)$ are complex conjugated to each other. 
We have also a $\mathbb{Z}_4$ grading $\tau$ (4-ality) on the set of irreps, such that $\tau(\overline{\lambda}) = -\tau(\lambda) \mod 4$ given by 
$\tau(\lambda_1,\lambda_2,\lambda_3) = \lambda_1+ 2 \lambda_2 + 3 \lambda_3 \mod 4$. 
Fusion coefficients are such that 
$\lambda \otimes \lambda' = \oplus_{\lambda''} {N}_{\lambda \lambda'}^{\lambda''} \; \lambda''$.  
Using Young tableaux techniques one obtains :   
\begin{eqnarray*}
(100)  \otimes (\lambda_1,\lambda_2,\lambda_3) &=& (\lambda_1 + 1 ,\lambda_2,\lambda_3) \oplus (\lambda_1 - 1,\lambda_2 + 1,\lambda_3) \oplus (\lambda_1,\lambda_2 - 1,\lambda_3+1) \\
{} & \oplus&  (\lambda_1,\lambda_2,\lambda_3-1) \\
(010) \otimes (\lambda_1,\lambda_2,\lambda_3)  &=& (\lambda_1,\lambda_2+1,\lambda_3) \oplus (\lambda_1+1,\lambda_2-1,\lambda_3+1) \oplus (\lambda_1+1,\lambda_2,\lambda_3-1) \\
{ } &\oplus& (\lambda_1-1,\lambda_2,\lambda_3+1) \oplus (\lambda_1-1,\lambda_2+1,\lambda_3-1) \oplus (\lambda_1,\lambda_2-1,\lambda_3) \\
(001) \otimes (\lambda_1,\lambda_2,\lambda_3)  &=& (\lambda_1,\lambda_2,\lambda_3+1) \oplus (\lambda_1+1,\lambda_2-1,\lambda_3) \oplus (\lambda_1,\lambda_2+1,\lambda_3-1)\\
{} &  \oplus & (\lambda_1-1,\lambda_2,\lambda_3)
\end{eqnarray*}
where one has to discard, from the right hand side,  those possible terms with  negative coordinates.
Fusion matrices $N_{(100)}$, $N_{(010)}$ and $N_{(001)}$ read from this
give the adjacency matrices of the graph\footnote{From now on we no longer make explicit the $SU(4)$ argument. } ${\mathcal A}_4={\mathcal A}_4 (SU(4))$  displayed in figure \ref{dimA4}. 
$N_{(100)}$ describes oriented edges from $\lambda$ to $\lambda'$, the arrows pointing in the direction of increasing 4-ality ($\tau(\lambda') = \tau(\lambda)+1 \mod 4$).
$N_{(001)}$ is the transposed of $N_{100}$, it  describes oriented edges from $\lambda'$ to  $\lambda$,  the arrows pointing in the direction of decreasing 4-ality. Both oriented edges of type $N_{(100)}$ and $N_{(001)}$ are drawn in red, without marking the arrows,  since the direction can be deduced from the 4-ality of vertices. The generator $N_{(010)}$ only connects vertices $\lambda$ and $\lambda'$ such that $\tau(\lambda') = \tau(\lambda)+2 \mod 4$, its edges are drawn in blue (bi-oriented).
Once these fusion matrices are known, the others can be determined from the {\sl truncated \/} recursion formulae of $SU(4)$ irreps, applied for increasing level $\ell$, up to $k$ ($2 \leq \ell \leq k=4$):
\begin{eqnarray}
N_{(\ell -p ,p-q,q)} &=& N_{(1,0,0)} \, N_{(\ell-p-1,p-q,q)} - N_{(\ell-p-2,p-q+1,q)} - N_{(\ell-p-1,p-q-1,q+1)} \\
{} &-&  N_{(\ell-p-1,p-q,q-1)}    \qquad \qquad \qquad \; \; \;\qquad \textrm{for } 0 \leq q \leq p \leq \ell - 1 \nonumber \\
N_{(0,\ell-q,q)} &=&  (N_{(q,\ell-q,0)})^{tr} \qquad \qquad \qquad \qquad \qquad \textrm{for } 1 \leq q \leq \ell  \nonumber \\
N_{(0,\ell,0)} &=& N_{(0,1,0)} \, N_{(0,\ell-1,0)} - N_{(1,\ell-2,1)} - N_{(0,\ell-2,0)}  \nonumber
\label{su4recrel}
\end{eqnarray} 
The fusion coefficients can also be obtained from $s$ and $t$ matrices by the Verlinde formula.

The quantum dimensions $\mu$ of vertices $\lambda = (\lambda_1,\lambda_2,\lambda_3)$ of $\mathcal{A}_4$ are obtained from the matrix $N_{100}$ by calculating the normalized eigenvector associated with the eigenvalue of maximal norm $\beta = [4]  = \sqrt{2 (2 + \sqrt{2})}$, where $[n] = \frac{q^n-q^{-n}}{q-q^{-1}}$, with $q=\exp(i \pi /\kappa)$. Here $\kappa=k+4=8$. They can also be obtained from the quantum Weyl formula applied to $SU(4)$:
$$
\mu_{\lambda} = \frac{[\lambda_1+1][\lambda_2+1][\lambda_3+1][\lambda_1+\lambda_2+\lambda_3+3][\lambda_1+\lambda_2+2][\lambda_2+\lambda_3+2]}{[2][2][3]}
$$
One finds
$\mu_{000} = [1] =1, \; \mu_{100} = \mu_{001} = [4] = \beta, \; \mu_{010} = [4][3]/[2] = 2 +\sqrt{2}$. Quantum dimensions form a one dimensional representation of the 
fusion algebra. 
The quantum mass (or quantum order) $|\mathcal{A}_4|$, which is the corresponding sum of squares, is $128(3+2\sqrt{2})$.

\paragraph{Toric matrices.}
 We determine the toric matrices $W_{z,0}$, of size $35\times 35$, by solving the modular splitting equation. 
For each choice of the pair $({\lambda , \mu})$ ($=35^2$ possibilities), we  define matrices $K_{\lambda\mu}$ by:
$K_{\lambda\mu} = N_{\lambda}\,\mathcal{M}\,N_{\mu}^{tr}$ and calculate them.
The modular splitting equation reads:  
\begin{equation}
K_{\lambda\mu} = \sum_{z=0}^{d_O-1} (W_{0,z})_{\lambda\mu} \, W_{z,0} \;.
\label{Kmse}
\end{equation}
It can be viewed as the linear expansion of the matrix $K_{\lambda\mu}$ over the set of toric matrices $W_{z,0}$, where the coefficients of this expansion are the non-negative integers $(W_{0,z})_{\lambda\mu}$ and where  $d_O=Tr(\mathcal{M}\mathcal{M}^{\dagger}) = 48$ is the dimension of the quantum symmetry algebra. These equations have to be solved for all possible values of $\lambda$ and $\mu$.  In other words,  we have a single equation for a huge tensor with $35^2 \times 35^2$ components but we prefer to view it as a family of  $35^2$ vectors $K_{\lambda\mu}$, each vector being itself a $35\times 35$ matrix. Using computer algebra techniques, one finds that matrices $K_{\lambda\mu}$ span a vector space of dimension $r = 33 < 48$. Therefore, the 48 toric matrices $W_{z,0}$ are not linearly independent.  This is not a surprise :  from the presence of a block $4\times 4$ in $Oc({\mathcal E}_4)$, one indeed expects the rank to be  $48-4^2+1=33$.
The toric matrices $W_{z,0}$ are obtained by using an algorithm explained in \cite{EstebanGil}. For each matrix $K_{\lambda\mu}$ we calculate its ``norm''  (abusive terminology) defined by $norm(K_{\lambda\mu}) =(K_{\lambda\mu})_{\overline{\lambda}\overline{\mu}}$, equal to the sum of the square of the coefficients appearing in the expansion of $K_{\lambda\mu}$ along the family (not a base) of toric matrices.
There is a subtlety here: it may happen that $W_{z_{1},0}=W_{z_{2},0}$ with $z_1 \neq z_2$, in that case one has to consider $W_{z_{1},0}$ and $W_{z_{2},0}$ as distinct when evaluating this sum.
\begin{itemize}
\item There are  8 linearly independent matrices $K_{\lambda\mu}$ of norm 1, each one therefore defines a toric matrix. 
\item There are 11 linearly independent matrices $K_{\lambda\mu}$ of norm 2. None of them is equal to the sum of two already determined toric matrices, and they cannot be written as a sum of a known toric matrix and a new one. Here and below, we rely on arguments using non negativity of the matrix elements.  These 11 matrices have elements that are multiple of 2. Dividing them by 2 we obtain in this way 11 new toric matrices appearing with multiplicity 2 in the family.
\item There are 8 linearly independent matrices $K_{\lambda\mu}$ of norm 3. Four of them are equal to the sum of three already determined toric matrices. Each of the other four can be written as a sum of an
an already determined toric matrix and a new one, whose coefficients are multiple of 2. 
Dividing them by 2 we obtain in this way 4 new toric matrices, with multiplicity 2.
At this stage we have obtained 8+11+4 = 23 linearly independent toric matrices, so we are still missing $10$ of them ($10 = 33-23$); however, counting multiplicities, we have $(8 \times 1)+(11 \times 2)+(4 \times 2) = 38 = 48-10$. We know that we must obtain 33 linearly independent toric matrices, but there are 48 toric matrices, so the last $10$, still missing, should be linearly independent.   
\item There are 5 linearly independent matrices $X=K_{\lambda\mu}$ of norm 4, none of them equal to the sum of two already determined toric matrices, and they have matrix elements that are multiple of 4. 
There are two writing possibilities giving length 4, either $X=2 W$, with a new toric matrix $W$ defined as $W=X/2$,  or $X = W+W'+W''+W'''$ with $W$ defined as $X/4$ and where primes refer to multiplicities. This last possibility is rejected since we already determined all toric matrices appearing with multiplicities. We obtain in this way 5 new toric matrices.
\item There are 6 linearly independent matrices $K_{\lambda\mu}$ of norm 5, but they can be written as sums of already determined toric matrices. 
\item There are 12 linearly independent matrices $X=K_{\lambda\mu}$ of norm 6, eight of them are sums of already determined toric matrices. The other four can be written as sums $X=W_{old} + W_{old}'+ 2 W$ where the first two terms are known toric matrices and the last term $W$ is new. In this way we obtain 4 more toric matrices.
\item There is nothing at norm 7. 
\item There are 3 linearly independent matrices $X=K_{\lambda\mu}$ of norm 8, but only one is not equal to the sum of already determined toric matrices; moreover its elements are multiple of $2$. It can be written as a sum $X=2 W_{old} + 2 W$ where the last term is new. 
\end{itemize}
We have therefore determined $33$  linearly independent toric matrices $W_{x,0}$, $15=11+4$ of them coming with multiplicity 2, so that the total number of toric matrices is indeed $(18\times1) + (15\times 2) = 48$. 
We can check that all other $K_{\lambda\mu}$ matrices can be expanded along the obtained family. We can also check that the modular splitting equation (\ref{Kmse}) is verified.

Ideally we would have liked to summarize the torus structure by displaying one toric matrix $W_{x,0}$ (a $35
\times 35$ matrix) for each vertex $x$ of its graph of quantum
symmetries (48 vertices), that we shall obtain later. 
This is obviously impossible in printed form, interested readers may
obtain this information from the authors.
The first matrix (which we knew already) describes  the modular invariant of the $SU(4)$ theory at level $4$ with boundary types specified by the exceptional graph ${\mathcal E}_4$;  
the other partition functions ($48-1$ of them) 
are not modular invariant (however they all commute with $s^{-1}.t.s$) and 
can be understood, in the interpretation of \cite{PetkovaZuber:Oc}, as  describing  the same BCFT theory but with defects labelled by $x$.
One possibility would be to give a table of the $48$ partition functions (only $33$ are distinct)
but to limit the size of this paper, we shall only give those
associated with three particular vertices called ambichiral and denoted $1\dot{\otimes} 1,2\dot{\otimes} 1$ and $9\dot{\otimes} 1$. 
Setting $u = (000 + 210 + 012 + 040), v = (101 + 400 + 121 + 004)$ and $w=(111)$, they read
\begin{eqnarray*}
\mathcal{Z}(1\dot{\otimes}1) &=& \mathcal{Z} = \vert u \vert^2 + \vert v \vert^2 + 4 \vert w \vert^2 \\
\mathcal{Z}(2\dot{\otimes}1)&=& u \,\overline{v} + \overline{u}\, v + 4 \vert w \vert^2 \\
\mathcal{Z}(9\dot{\otimes}1)&=& 2 (u+v)\overline{w} + 2\, w (\overline{u} + \overline{v})
\end{eqnarray*}

\begin{figure}
\centerline{\scalebox{0.75}{\includegraphics{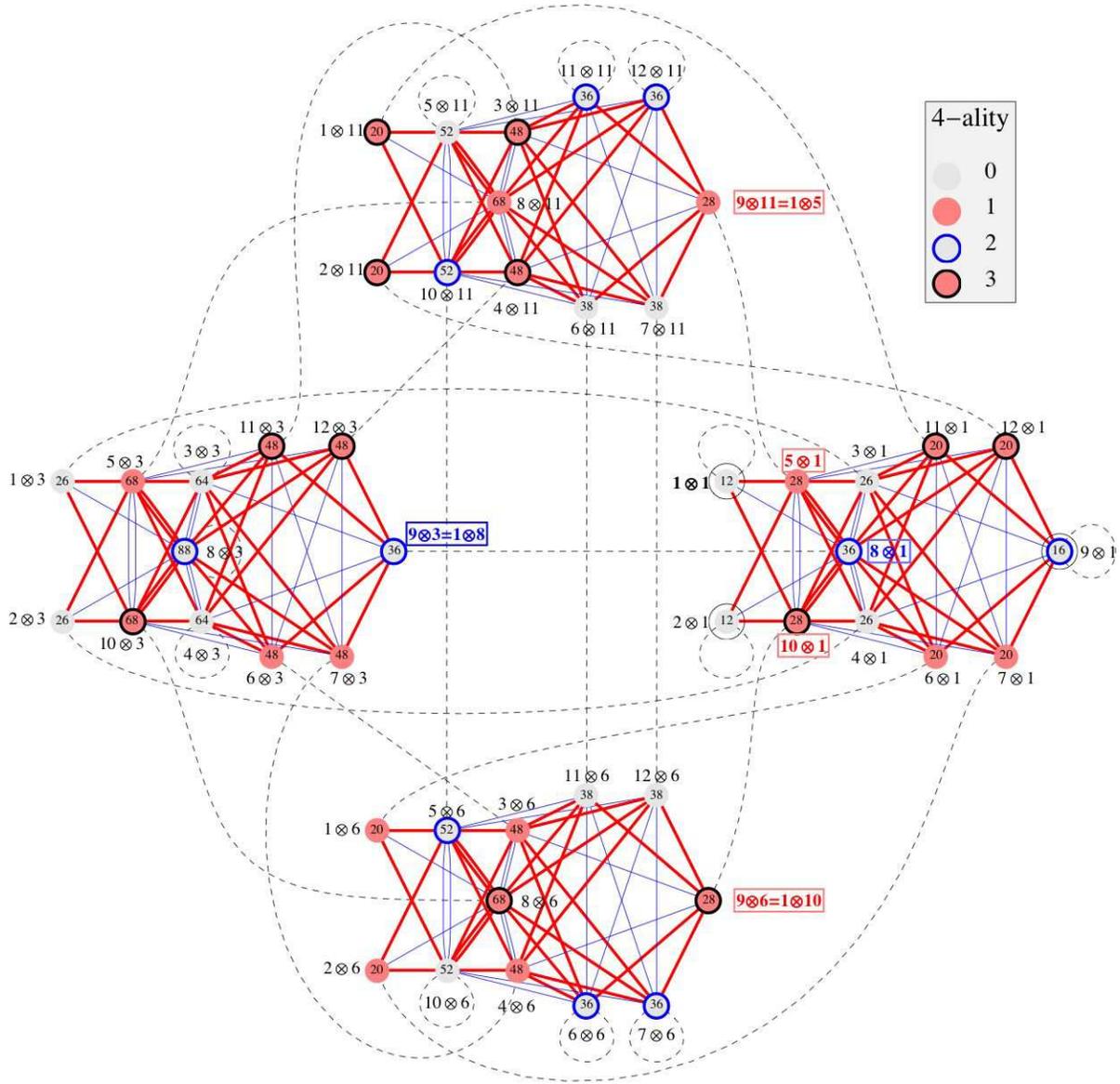}}} 
\caption{The Ocneanu graph of quantum symmetries $Oc(\mathcal{E}_4)$. The identity is $1 \otimes 1$. The left generators are $5 \otimes 1, 8 \otimes 1$ and $10 \otimes 1$, the right generators are $1 \otimes 5, 1 \otimes 8$ and $1 \otimes 10$. Multiplication by $5 \otimes 1$ (resp. $10 \otimes 1$) is encoded by oriented red edges (thick lines), in the direction of increasing (resp. decreasing) 4-ality. Multiplication by $8 \otimes 1$ is encoded by unoriented blue egdes (thin lines). Dashed lines relate chiral conjugated vertices. The three ambichiral vertices are circled on the graph.  Numbers on the vertices give the dimension of the simple blocks $x$ of the bialgebra $\mathcal{B}({\mathcal E}_4)$ for the multiplication on its dual.}
\label{Oc-E4}
\end{figure}

\paragraph{Chiral generators for the algebra of quantum symmetries.}
Using toric matrices $W_{x,0}$ and fusion matrices $N_{\lambda}$, we calculate, for every choice of $x,\lambda,\mu$, the matrices $K_{\lambda\mu}^x = N_{\lambda}W_{x,0}N_{\mu}^{tr}$. When $x=0$ we recover the matrices $K_{\lambda\mu}$ used previously, since $W_{0,0} = {\mathcal M}$.
We then decompose them on the family (not a base) of toric matrices $W_{z,0}$:
\begin{equation}
K_{\lambda\mu}^x = \sum_z (W_{x,z})_{\lambda\mu} W_{z,0} \;.
\label{klmx}
\end{equation}
The coefficients of this expansion, that we want to determine, define ``toric matrices with two twists''  $(W_{x,z})_{\lambda\mu} = (V_{\lambda\mu})_{xz}$. 
Since the $W_{z,0}$ are not linearly independent on $\CC$, the decomposition (\ref{klmx}) is not unique, and there are some undetermined coefficients. Imposing that they should be non-negative integers fixes some of them or allows to obtain relations between them,  
one can also use the intertwining property $V_{\lambda\mu} ( N_\lambda \otimes N_\mu ) = (N_\mu \otimes N_\lambda )  V_{\lambda\mu} $. 
The group $SU(4)$ has three fundamental irreducible representations $f$, therefore the graph of quantum symmetries has $6$ (chiral) generators, three left $f^L$ and 
three right $f^R$. Multiplication by these generators is encoded by the quantum symmetry matrices $O_{f^L} = V_{f0}$ and $O_{f^R}=V_{0f}$.
Choosing an appropriate order  on the set of indices $z$, we obtain the following structure for the left chiral generators  $V_{100,000}$ and $V_{010,000}$, (the last chiral left generator is $V_{001,000} = V_{100,000}^{tr}$).  
{\small
\begin{equation}
V_{100,000} = \left(
\begin{array}{cccccc}
F_{100} & . & . & . \\
. & F_{100}  & . & . \\
. & . & F_{100}  & . \\
. & . & . & F_{100}
\end{array}
\right)
\qquad 
V_{010,000} = \left(
\begin{array}{cccccc}
F_{010} & . & . & . \\
. & F_{010}  & . & . \\
. & . & F_{010}  & . \\
. & . & . & F_{010}
\end{array}
\right)
\end{equation}
}
Here  $F_{100}$ and $F_{010}$ denote explicit $12\times 12$ matrices; they still have undetermined coefficients reflecting the existence of classical $Z_2$ symmetries  but they can be determined once an ordering has been chosen (see later). 
The right chiral generators are also essentially known at this step: to fix the last coefficients,  one uses the fact that right generators are conjugated from the left ones by an appropriate permutation matrix $P$, the chiral conjugation,  acting on the 48 vertices of the graph: $O_{f^R} = P\,O_{f^L}\,P^{-1}$).
From the knowledge of the six chiral generators, we can draw the two chiral subgraphs making the
Ocneanu graph of quantum symmetries: see figure \ref{Oc-E4}. 
Actually, it is enough to draw the left graph, which describes the multiplication of an arbitrary vertex by a chiral left generator:  red edges, oriented in the direction of increasing or decreasing 4-ality, or non-oriented blue edges. On the graph, chiral conjugated vertices are related by a dashed line. Multiplication of a vertex $x$ by the chiral generator $f^R$ is obtained as follows: we start from $x$, follow the dashed lines to find its chiral conjugate vertex $y$, then use the multiplication by $f^L$ and finally pull back using the dashed lines to obtain the result. 

Remark. The six chiral generators are ``generators'' in the sense that all possible quantum symmetries $O_x$ appear on the right hand side when we multiply these generators together in all possible ways. However, because of existence of classical symmetries (more about it later), the chiral generators only generate a subalgebra of $Oc({\mathcal E}_4)$,  of dimension $33$ : 18 quantum symmetries of type $O_x$ and 15 ``composites''  (sums of two) of  type  $O_{x'} + O_{x''}$. 
Because of this compositeness, the algebra spanned by the left ``generators'' is only a commutative subalgebra of dimension $9=12-3$ of the non commutative left chiral algebra (of dimension $12$), same thing for the right part. Non commutativity of $Oc({\mathcal E}_4)$ is expected from the presence of a coefficient $4>1$ on the diagonal of  the modular invariant, and non commutativity of  the two chiral subalgebras is expected from the fact that they should nevertheless commute with each other.
Because of the known properties of quantum symmetries, we shall actually be able to exhibit a matrix realization for {\sl all \/} generators, \ie to lift the degeneracy between sums of two, but first we have to discuss the graph ${\mathcal E}_4$ itself,  whose structure is encoded by adjacency matrices $F_{100}$,  $F_{010}$ and $F_{001}$.

\begin{figure}
\centerline{\scalebox{0.5}{\includegraphics{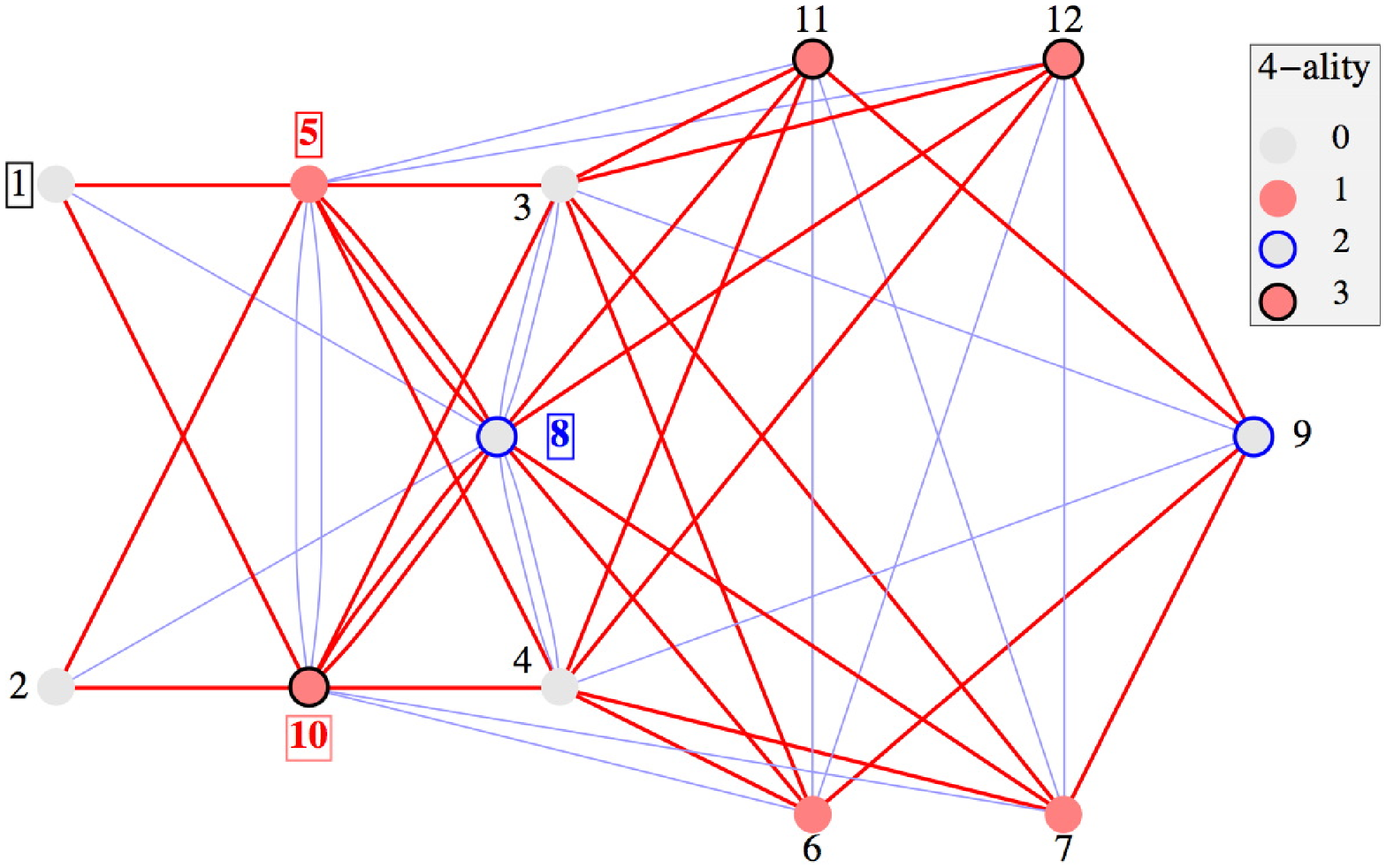}}} 
\caption{The $\mathcal{E}_4$ quantum graph. Generators are called 5, 8 and 10. Identity is 1. Vertices 5 and 10 are conjugated, 8 is self-conjugated. Multiplication by 5 (resp. 10) is encoded by red edges (thick lines) in the direction of increasing (resp. decreasing) 4-ality, multiplication by 8 is encoded by unoriented blue edges (thin lines).}
\label{gr-E4}
\end{figure}

\paragraph{The  exceptional quantum graph ${\mathcal E}_4$.}
It  appears in its graph of quantum symmetries.
The $F_{100}=F_{001}^{tr}$ and $F_{010}$ matrices are the adjacency matrices of the graph $\mathcal{E}_4$, with $12$ vertices,  displayed on figure \ref{gr-E4}. This graph possesses $\mathbb{Z}_2$-symmetries corresponding to the permutation of vertices $3-4$, $6-7$ and $11-12$. The undetermined coefficients of the adjacency matrix reflect this symmetry and are simply determined once an ordering has been chosen for the vertices (something similar happens for the $D_{even}$ series of the $su(2)$ family). 
We can define a coloring $\tau$ (4-ality) on this quantum graph. 
$F_{100}$ (resp. $F_{001}$) corresponds to edges in red, pointing in the direction of increasing (decreasing) 4-ality. $F_{010}$ corresponds to the blue edges (bi-oriented).
 The quantum dimensions $\mu_a$ of the vertices of $\mathcal{E}_4$ are obtained from the adjacency matrix of $F_{100}$ by calculating the normalized eigenvector associated with the eigenvalue  of maximal norm $\beta =  \sqrt{2 (2 + \sqrt{2})}$. One finds
 $\mu_1 = \mu_2 = 1, \mu_3=\mu_4 = 1 + \sqrt{2}$,  $\mu_5 = \mu_{10} = \sqrt{2 (2 + \sqrt{2})}$, $\mu_6=\mu_7 = \mu_{11} = \mu_{12} = \sqrt{2 + \sqrt{2}} $,  $\mu_8 = 2 + \sqrt{2}$, $\mu_9 = \sqrt{2}$. The quantum mass (or quantum order), the corresponding sum of squares, is $|\mathcal{E}_4| =16(2+\sqrt{2})$.

\paragraph{${\mathcal E}_4$ as a module over ${\mathcal A}_4$.}
The vector space of $\mathcal{E}_4$ is a module over the graph algebra (fusion algebra) of $\mathcal{A}_4$, encoded by the annular matrices $F_n$
\begin{equation}
\mathcal{A}_4 \times \mathcal{E}_4 \rightarrow \mathcal{E}_4 : \quad \lambda_n \cdot a =
 \sum_{b} (F_{n})_{ab}\;b \qquad \qquad \lambda_n \in \mathcal{A}_4\;, \quad a,b \in \mathcal{E}_4 \;.
\end{equation}
The $F_{n}$ matrices are determined from the generators $F_{100},F_{010}$ and $F_{001}$ by the same recursion relation as for fusion matrices (\ref{su4recrel}). 
Conjugation compatible with the $\mathcal{A}_4$ action can be defined on $\mathcal{E}_4$. It reads $\overline{1}=1, \overline{2}=2, (\overline{3+4})=(3+4), \overline{5}=10, (\overline{6+7})=(11+12), \overline{8}=8, \overline{9}=9$, but it is not entirely determined at this level, since there is still an ambiguity (solved later) on the definition of conjugation for members of the doublets.  Notice that fundamental matrices (for instance $F_{100}$) contain, in this case, elements bigger than 1. However, the ``rigidity condition'' $(F_n)_{ab}= (F_{\overline{n}})_{ba}$ holds, so that this example is indeed an higher analogue of the $ADE$ graphs, not an higher analogue of the non simply laced cases.

\paragraph{The first multiplicative structure of the quantum groupo\"\i d.}
Now that annular matrices $F_\lambda$ are known, we can calculate the dimensions $d_\lambda$ of blocks of the quantum groupo\"\i d ${\mathcal B}$ for its first multiplicative structure. They are given by 
$d_\lambda = \sum_{a,b} (F_\lambda)_{ab}$.
These numbers appear in figure \ref{dimA4}. Notice that the dimension of the horizontal vector space (essential paths) is $\sum d_\lambda = 1568$, and $dim({\mathcal B}) = \sum d_\lambda^2 = 86816$.

\paragraph{Self-fusion on $\mathcal{E}_4$.}  This graph (displayed on figure \ref{gr-E4}) has self-fusion: the vector space spanned by its $12$ vertices has an associative algebra structure, with non-negative structure constants, and it is compatible with the action of $\mathcal{A}_4$. 
There could be, a priori,  several possibilities, however, because of the expected non commutativity of Oc, and from the fact that the span of left chiral generators commute with the span of right chiral generators, we are looking for a non commutative structure for the self-fusion of ${\mathcal E}_4$. Up to isomorphism there is only one non commutative solution that we now describe.
$1$ is the unity and the three generators are $5$, $8$ and $10$. ($10$ is conjugated to $5$, $8$ being self conjugated). The graph algebra matrices encoding this multiplication are given by $G_1 = \munite, G_5 = F_{100}, G_8 = F_{010}, G_{10}=G_5^{tr}, G_2= 2\,(G_5\,G_{10})-G_1-G_8\,G_8, G_9=(1/2)(G_8 \, (G_5\,G_{10}-G_1-G_2))-2G_8$ and $(G_3+G_4)=G_5\,G_{10}-G_1-G_2 \, , \, (G_6+G_7)=G_8\,G_{10}-G_5-G_5 \, , \, (G_{11}+G_{12})=G_5\,G_{8}-G_{10}-G_{10}$.
Due to the symmetries of the graph $\mathcal{E}_4$, the knowledge of the multiplication by generators $5$, $8$ and $10$ is not sufficient to reconstruct the whole multiplication table, even after imposing that structure coefficients should be non-negative integers, since we cannot split the doublets $(3+4)$, $(6+7)$, $(11+12)$. Imposing the additional condition $G_{\overline{a}} = G_a^{tr}$ such that multiplication by $\overline{a}$ is obtained by reversing the arrows in the graph of multiplication by $a$ leads, up to isomorphism, to a unique solution. It fixes the conjugation on the graph to be\footnote{The other solution with $\overline{3}=4$ leads to non-integer structure coefficients and it is discarded.} $\overline{3}=3, \overline{4}=4$, and $\overline{6}=11, \overline{7}=12$, and the obtained multiplication structure appears to be non commutative. We give below the table of multiplication for vertices appearing in doublets. 
\begin{table}[hhh] 
$$ 
\begin{array}{|c||c|c|c|c|c|c|} 
\hline 
{\nearrow}& 3 & 4  & 6 & 7 & 11 & 12  \\ 
\hline 
\hline 
3 & 1+3+4  & 2+3+4 & 5+7  & 5+6 & 10+11 & 10+12  \\ 
\hline
4 & 2+3+4  & 1+3+4 & 5+6  & 5+7  & 10+12 & 10+11  \\ 
\hline
6 & 5+6  & 5+7 & 8  & 8 & 1+4 & 2+3  \\ 
\hline
7 & 5+7  & 5+6 & 8  & 8    & 2+3 & 1+4  \\ 
\hline
11 & 10+12  & 10+11 & 1+3  & 2+4    & 8 & 8  \\ 
\hline
12 & 10+11  & 10+12 & 2+4  & 1+3    & 8 & 8  \\  
\hline 
\end{array} 
$$ 
\caption{Multiplication table for vertices that are members of doublets in $\mathcal{E}_4$.} 
\end{table} 
Call $t$ the twist exchanging the two members of the same doublet $t(3)=4, t(4)=3$, $t(6)=7, t(7)=6$, $t(11)=12, t(12)=11$ and leaving the others invariant $t(i)=i$. 
Using the multiplication table of $\mathcal{E}_4$ we can check that $t$ is an involution such that $t(a\cdot b) = t(b) \cdot t(a)$. The non-commutativity can be seen from:
$$
a \cdot b = t(b) \cdot a \, \not = b \cdot a  \qquad \qquad \text{for } a,b \notin \text{the same doublet}
$$

\paragraph{Realization of  $Oc(\mathcal{E}_4)$.}
Notice that each chiral subgraph has itself four components, as it is clear from the block diagonal structure of the generators.  The first component describe a (chiral) subalgebra, and the other three are modules over it. Ambichiral generators span the intersection of left and right chiral subalgebras,  this subalgebra has dimension $3$ and the associated vertices belong to the first component of the full graph; they are self-dual generators (both left and right), of course, but there are altogether $12$ self-dual generators : three for each component.

We can use the previous explicit realization of the graph algebra of ${\mathcal E}_4$  to build explicitly all generators of its algebra of quantum symmetries $Oc(\mathcal{E}_4)$. This algebra can be realized as
\begin{equation}
Oc(\mathcal{E}_4) = \mathcal{E}_4 \otimes_J \mathcal{E}_4 = \mathcal{E}_4 \dot{\otimes} \mathcal{E}_4 \qquad \qquad
\textrm{with } a \dot{\otimes} (b \cdot c) = (a \cdot b) \dot{\otimes} c \qquad \textrm{for } b \in J \,,
\end{equation} 
where $J$ is the modular subalgebra of the graph algebra $\mathcal{E}_4$, spanned by $\{1,2,9\}$,  isomorphic with the ambichiral subalgebra of $Oc(\mathcal{E}_4)$ defined later; it has particular modular properties\footnote{Using induction, one sees that its three vertices are associated with the three blocks of the modular invariant.}
(see \cite{GilRobert2,Gil:Thesis}).  
The left chiral subalgebra $L$ is spanned by elements of the form $a \dot{\otimes}  1$ and 
the right chiral subalgebra $R$ is spanned by elements of the form $1 \dot{\otimes}  a$, where $a$ are the basis elements of $\mathcal{E}_4$ associated with the vertices of its graph.
All elements $a$ of $\mathcal{E}_4$ can be written in the form $a=c \, j = j \, c$ for $c \in \{1, 3, 6, 11 \}$ and $j \in J$;  for example we have $4=3 \cdot 2$, $8= 3 \cdot 9$.  The base $\{ 1 \dot{\otimes} a \}$ of the right chiral subalgebra $R$ can therefore be written $ \{ j \dot{\otimes} 1, j \dot{\otimes} 3, j \dot{\otimes} 6, j \dot{\otimes} 11 \}$ with $a \in \mathcal{E}_4, j \in J$. 
The left chiral fundamental generators are  $5 \dot{\otimes} 1$, $8 \dot{\otimes} 1$ and $10 \dot{\otimes} 1$.  The right chiral fundamental generators are  $1 \dot{\otimes} 5 = 9 \dot{\otimes} 11$, $1 \dot{\otimes} 8 = 9 \dot{\otimes} 3$ and $1 \dot{\otimes} 10 =  9 \dot{\otimes} 6$.
Their intersection defines the ambichiral subalgebra of $Oc(\mathcal{E}_4)$ (also called $J$) spanned by 
$\{ 1 \dot{\otimes} 1, 2 \dot{\otimes} 1, 9 \dot{\otimes} 1 \} $. 

A natural basis of $Oc(\mathcal{E}_4)$ is given by elements $a \dot{\otimes} b$, but they can be written in terms of  $a \dot{\otimes} 1$, $a \dot{\otimes} 3$, $a \dot{\otimes} 6$, $a \dot{\otimes} 11$. 
The identifications in $Oc(\mathcal{E}_4)$ are given by: 
\begin{equation}
\begin{array}{rclcrcl} 
a \dot{\otimes} 2 &=& (2 \cdot a)  \dot{\otimes} 1  &\qquad& a \dot{\otimes} 9 &=& (9 \cdot a) \otimes 1   \\ 
a \dot{\otimes} 4 &=& (2 \cdot a)  \dot{\otimes} 3  &\qquad& a \dot{\otimes} 8 &=& (9 \cdot a) \otimes 3   \\ 
a \dot{\otimes} 7 &=& (2 \cdot a)  \dot{\otimes} 6  &\qquad& a \dot{\otimes} 10 &=& (9 \cdot a) \otimes 6   \\ 
a \dot{\otimes} 12 &=& (2 \cdot a)  \dot{\otimes} 11  &\qquad& a \dot{\otimes} 5 &=& (9 \cdot a) \otimes 11  
\end{array}
\label{ide4} 
\end{equation}
The chiral conjugation is obtained by $(a \dot{\otimes} b)^C = b \dot{\otimes} a$. 
Complex conjugation in $Oc(\mathcal{E}_4)$ is defined by $(a \dot{\otimes} b)^* = \overline{a} \dot{\otimes} \overline{b}$. 
Multiplication in $Oc(\mathcal{E}_4)$ is obtained from the multiplication of $\mathcal{E}_4$:
$$
(a \dot{\otimes} b) \, \cdot \, (a' \dot{\otimes} b') = (a\cdot a') \, \dot{\otimes} \, (b \cdot b') \;, 
$$
together with the identifications (\ref{ide4}), and it is encoded by the quantum symmetries matrices $O_x$. Choosing the basis ordering $\{ a \dot{\otimes } 1 ,a \dot{\otimes } 3, a \dot{\otimes } 6, a \dot{\otimes } 11 \}$ these matrices read:
{\small
$$
O_{x=a\dot{\otimes} 1} = 
\left(
\begin{array}{cccc}
G_a & . & . & . \\
.   & G_a & . & .  \\
.   & . & G_a & .  \\
.   & . & . & G_a  
\end{array}
\right)
\qquad
O_{x=a\dot{\otimes} 3} = 
\left(
\begin{array}{cccc}
. & G_a & . & . \\
G_a & G_a \,(G_1+G_2) & . & .  \\
. & . & G_2 \, G_a & G_9 \, G_a  \\
. & . & G_9 \, G_a & G_a  
\end{array}
\right)
$$

$$
O_{x=a\dot{\otimes} 6} = 
\left(
\begin{array}{cccc}
. & . & G_a & . \\
. & . & G_a & G_9 \, G_a  \\
. & G_9 \, G_a & . & .  \\
G_a & G_2 \, G_a & . & .  
\end{array}
\right)
\qquad
O_{x=a\dot{\otimes} 11} = 
\left(
\begin{array}{cccc}
. & . & . & G_a \\
. & . & G_9 \, G_a & G_2 \, G_a  \\
G_a & G_a & . & .  \\
. & G_9 \, G_a & . & .  
\end{array}
\right)
$$}
One can check the following quantum mass rule formula:
$$
|{Oc}| = |\mathcal{A}_4| \qquad \qquad \text{with } \qquad |{Oc}| = \frac{|\mathcal{E}_4| \, |\mathcal{E}_4|}{|J|}
$$
Indeed, $|\mathcal{A}|=128(3+2\sqrt{2})$, $|\mathcal{E}|=16(2+\sqrt{2})$ and $|J| = (1,1,\sqrt{2})(1,1,\sqrt{2}) = 4$.  
Now that the graph of ${\mathcal E}_4$ itself is known, as well as the annular matrices $F_{\lambda}$ and the realization of $Ox$ generators as tensor products, we can check that our determination of toric matrices $W_{x,0}$ was indeed correct: we calculate the rectangular essential matrices $E_a$ defined by  $(E_a)_{nb} = (F_n)_{ab}$ and the so-called ``reduced essential matrices'' $E_a^{red}$ obtained from the $(E_a)$ by keeping only the columns relative to the modular subalgebra $J$ (replace all other coefficients by $0$) \cite{Coque:Qtetra}, and finally check the identity  $W_{x,0} =  E_a . E_b^{red}$ whenever $x = a \dot{\otimes} b$. More generally, once the multiplication in $Oc(\mathcal{E}_4)$ is known, we can recover the toric matrices with two twists thank's to the relation $W_{x,y} = W_{xy,0}$.  
From the knowledge of the intertwiner matrix $E_0$, one can also obtain the simple summands of the associated Frobenius algebra   (they play the role of quantum Klein invariants): 
${\mathcal F}= \lambda_{000} \oplus \lambda_{210} \oplus \lambda_{012} \oplus \lambda_{040}.$
In this particular case, they could also be obtained from the structure of the identity block of the modular invariant or from the first column of the essential matrix relative to the unit vertex of ${\mathcal E}_4$.

\paragraph{${\mathcal E}_4$ as a module over $Oc({\mathcal E}_4)$.}
The vector space of $\mathcal{E}_4$ is a module over the algebra of quantum symmetries $Oc(\mathcal{E}_4)$, the action being defined by:
\begin{equation}
Oc \times \mathcal{E}_4 \rightarrow \mathcal{E}_4 \qquad \qquad \qquad
(a \dot{\otimes} b) \, \cdot \, c  \doteq  a \, \cdot \, c \, \cdot \, t(b) \;.
\end{equation} 
We can check the module property $\left[ (a_1 \dot{\otimes} b_1 ) \cdot (a_2 \dot{\otimes} b_2)\right] \cdot c = (a_1 \dot{\otimes} b_1 ) \cdot \left[ (a_2 \dot{\otimes} b_2) \cdot c \right] $
using the fact that $t$ is an involution $(b_1 \cdot b_2)^t = b_2^t \cdot b_1^t$. 
The dual annular matrices $S_x$ encoding this action $x \cdot c = \sum_d (S_x)_{cd} \, d$ are given by:
\begin{equation}
S_{x = a \otimes_J b} \doteq G_a \, G_{b^{t}}' \;,
\end{equation}
where the $G_b'$ matrices are defined by $a \cdot b = \sum_{c} (G_{b}')_{ac} \, c$ given by $(G_b')_{ac} = (G_{a})_{bc}$.

\paragraph{The second  multiplicative structure of the quantum groupo\"\i d.}
Now that dual annular matrices $S_x$ are known, we can calculate the dimensions $d_x$ of blocks of the quantum groupo\"\i d ${\mathcal B}$ for its second multiplicative structure (the multiplication on the dual $\widehat{{\mathcal B}}$). These numbers $d_x=\sum_{a,b} (S_x)_{ab}$ appear on figure \ref{Oc-E4}. The dimension of the vertical vector space  is $\sum d_x = 1864$, and $dim(\widehat{{\mathcal B}}) = \sum d_x^2 = 86816$, equal to $dim({\mathcal B}) = \sum d_\lambda^2$ as it should. The ``linear sum rule'' does not hold (total horizontal and vertical dimensions are not equal).

\paragraph{Matrix units and block diagonalization of ${\mathcal E}_4$
and of $Oc({\mathcal E}_4)$.}
With $u=\sqrt 2$ , $v=u+1$ and  $w=u+2$, consider the $8 \times 12$ matrix $X$ defined by the  table
{\small
$$
\begin{array}{llllllllllll}
 {\sqrt w} & {\sqrt w} & v {\sqrt w} & v {\sqrt w} & 2 v & u+2 & u+2 & w^{3/2} & u {\sqrt w} & 2 v & u+2 & u+2 \\
 u+2 & u+2 & -u & -u & 2 {\sqrt w} & -u {\sqrt w} & -u {\sqrt w} & 2 & -2 v & 2 {\sqrt w} & -u {\sqrt w} & -u {\sqrt w} \\
 {\sqrt w} & {\sqrt w} & v {\sqrt w} & v {\sqrt w} & 2 i v & i w & i w & -w^{3/2} & -u {\sqrt w} & -2 i v & -i w & -i w \\
 u+2 & u+2 & -u & -u & 2 i {\sqrt w} & -i u {\sqrt w} & -i u {\sqrt w} & -2 & 2 v & -2 i {\sqrt w} & i u {\sqrt w} & i u {\sqrt w} \\
 u+2 & u+2 & -u & -u & -2 i {\sqrt w} & i u {\sqrt w} & i u {\sqrt w} & -2 & 2 v & 2 i {\sqrt w} & -i u {\sqrt w} & -i u {\sqrt w} \\
 {\sqrt w} & {\sqrt w} & v {\sqrt w} & v {\sqrt w} & -2 i v & -i w & -i w & -w^{3/2} & -u {\sqrt w} & 2 i v & i w & i w \\
 u+2 & u+2 & -u & -u & -2 {\sqrt w} & u {\sqrt w} & u {\sqrt w} & 2 & -2 v & -2 {\sqrt w} & u {\sqrt w} & u {\sqrt w} \\
 {\sqrt w} & {\sqrt w} & v {\sqrt w} & v {\sqrt w} & -2 v & -u-2 & -u-2 & w^{3/2} & u {\sqrt w} & -2 v & -u-2 & -u-2
\end{array}
$$
}
and, for $s \in \{1,\ldots 8 \}$, the  coefficients $n(s)$ 
defined by $n(1)=n(3)=n(6)=n(8)= 16 (u+2)^{3/2}$,  and $n(2)=n(4)=n(5)=n(7)=32$.

For $s \in \{1, \ldots, 8 \}$ , take the eight  $12 \times 12$
matrices defined by 
$$\mu[s] =\mu[s,s] =\frac{1}{n(s)}  \sum_{q=1}^{q=12} \, X(s,q) G_q$$
and finally  the four matrices
{\small
\begin{eqnarray*}
\mu[9,9] =\mu[11,11]=1/4 (G_1-G_2+G_3-G_4) & \qquad
\mu[9,10] =\mu[11,12]=1/(2 u) ( G_{11}- G_{12}) \\
\mu[10,10] =\mu[12,12]=1/4 (G_1-G_2-G_3+G_4)
 &  \qquad
\mu[10,9] = \mu[12,11]=1/(2 u)( G_6- G_7)
\end{eqnarray*}
}
One can then check that the elements  $\mu[a,b]$ of the algebra $\mathcal{E}_4$  spanned by the $G_a$'s are matrix units of its commutant.
Indeed $\mu[a,a] \, \mu[a,a] = \mu[a,a]$, for
$a \in \{1,\ldots, 8\}$, and $\mu[a,b]\,  \mu[b,c] = \mu[a,c]$, when $a,b,c$ belong to $\{9,10\}$ or to $\{11,12\}$.
In other words, one can find  a $12 \times 12$ unitary matrix $U$
such that $U. \mu[a,b] . U^{-1}$ are elementary matrices belonging to the algebra spanned by the $G_a$'s: they contain
only a $1$ in position $(a,b)$.  The first $8$ are diagonal, and the
last $8$ generate an algebra isomorphic with $M(2,\CC)$.
Conversely, any linear combination of graph matrices $G_a$ can be
brought to this block diagonal form.
The graph algebra of ${\mathcal E}_4$ is therefore isomorphic, on the
complex field $\CC$,  with the algebra  $\bigoplus_{x=1}^{x=8}  \CC_x \oplus M(2,\CC)$.
The eight one-dimensional blocks have multiplicty one, whereas the block $M(2,\CC)$ has multiplicity $2$.
From the fact that $Oc({\mathcal E}_4)$ can be written as a tensor product over the subalgebra $J$ of its left and right chiral subalgebras, both isomorphic with $\mathcal{E}_4$,  one can block diagonalize it and show that it is isomorphic with the algebra $\bigoplus_{x=1}^{x=32}  \CC_x \oplus M(4,\CC)$. The thirty-two one-dimensional blocks  have multiplicity one, 
whereas the block  $M(4,\CC)$ has multiplicity $4$. This is in agreement with what was expected from the structure of the modular invariant.

\paragraph{The graph ${\mathcal E}_4$ : concluding comments.}
The purpose of the last section, devoted to the exceptional quantum graph $\mathcal{E}_4 (SU(4))$, 
 was to illustrate in a particular example, and using conformal embeddings,  the general features of the algebraic concepts described in the first part. 
The conformal embedding itself was known twenty years ago  \cite{SchellekensWarner}, the corresponding modular invariant already appears in \cite{AltschulerBauerItzykson, SchellekensYankielowicz} and the associated quantum graph was first found in \cite{PetkovaZuberNP1995}. From a study of the modular splitting equation, the later was recovered in  \cite{Ocneanu:Bariloche} and  presented in Bariloche (the existence of a self-connection satisfying the required compatibility  equations was also checked). However,  because of the heaviness of the involved calculations, only the first line of the modular invariant was used  in \cite{Ocneanu:Bariloche} to achieve this goal:  indeed, one can see that this provides enough information to obtain the graph itself.
To our knowledge, the full modular splitting system had not been solved, the full torus structure had not been obtained, and the graph of quantum symmetries was not known; this is what we did. We have also determined the whole (non-commutative) multiplicative structure of $Oc({\mathcal E}_4)$, not only a description of the action of its chiral generators. We obtained this information without having to rely on a separate study of the bialgebra of double triangles, something which would be intractable anyway in the $SU(4)$ situation. To our knowledge, this also is new.

\section*{Afterword}

Quantum graphs may have self - fusion or not:  as mentioned before, in the case of $SU(4)$, besides several infinite series of graphs with self-fusion, there are also  three exceptional quantum graphs with self-fusion, at levels $4$, $6$ and $8$, and the three corresponding modular invariants can be obtained from appropriate conformal embeddings\footnote{The fact that the last one can also be obtained in this way \cite{AldazabalEtAl} is a property that was apparently forgotten.}.  To conclude, we give the modular invariant partition functions 
  ${\mathcal Z}$ corresponding to modular invariant matrices ${\mathcal M}$ of respective sizes $35 \times 35$, $84 \times 84$ and $165 \times 165$, for the graphs ${\mathcal E}_4 (SU(4))$, ${\mathcal E}_6 (SU(4))$  and ${\mathcal E}_8(SU(4))$.
{\small
\begin{eqnarray*}
\mathcal{Z}(\mathcal{E}_4) &=& \vert 000 + 210 + 012 + 040 \vert^2 + \vert 101 + 400 + 121 + 004 \vert^2 + 4 \vert 111 \vert^2 \\
{}&{}&{}\\
\mathcal{Z}(\mathcal{E}_6) &=& \vert 006 + 022 + 220 + 600 \vert^2 + \vert 012 + 230 + 303 \vert^2 + \vert 002 + 212 + 240 \vert^2 + \vert 030 + 103 + 321\vert^2\\
{ }                        &+& \vert 032 + 210 + 303  \vert^2      + \vert 030 + 123 + 301 \vert^2 + \vert 042 + 200 + 212 \vert^2 + \vert 000 + 060 + 202 + 222 \vert^2\\
{ }                        &+& \vert 004 + 121 + 420  \vert^2 + \vert 024 + 121 + 400\vert^2 \\
{}&{}&{}\\
\mathcal{Z}(\mathcal{E}_8) &=& \vert 000 + 121 + 141 + 412 + 214 + 800 + 080 + 008 \vert^2 + 2 \,  \vert 311+113+331+133 \vert^2 \\
{ }                        &+& \vert 020 +230+032+060+303+602+323+206\vert^2  
\end{eqnarray*}}


\begin{thebibliography}{35}


\bibitem{AldazabalEtAl} Aldazabal G Allekote I  Font A and Nu{\~ n}ez C, 
1992 {N=2 Coset compactifications with nondiagonal invariants}  {\it Int Jour of Mod Phys}, {\bf A, Vol 7, No 25} pp  6273-6297

\bibitem{AltschulerBauerItzykson}   Altschuler D  Bauer M  and   Itzykson C 1990    {The Branching Rules of Conformal Embeddings}   {\it Commun  Math  Phys}  {\bf 132}  pp 349-364 

\bibitem{BaisBouwknegt}  Bais F  and  Bouwknegt P  1987   {A classification of subgroup truncations of the bosonic string}   {\it Nucl  Phys}  {\bf B279} p 561  

\bibitem{Evans-I}   B\"ockenhauer J and   Evans D  2000   {Modular invariants from subfactors: Type I coupling matrices and intermediate subfactors}  {\it Commun  Math  Phys}   {\bf 213 no 2}  pp 267-289  (preprint math OA/9911239) 

\bibitem{Evans-II}  B\"ockenhauer J  and   Evans D 1999  {Modular invariants  graphs and $\alpha$ induction for nets of subfactors II }
{\it Commun  Math  Phys} {\bf  200}  pp 57-103   

\bibitem{Evans-Kawahigashi}   B\"ockenhauer  J Evans D   and   Kawahigashi Y 2000  {Chiral structure of modular invariants for subfactors}
{\it Commun  Math  Phys} {\bf  210}  pp 733-784  

\bibitem{BohmSzlachanyi}   B\"ohm  G and  Szlach\'anyi K   1996  {A coassociative $C^*$-quantum group with non-integral dimensions}     {\it Lett  Math  Phys}   {\bf 38   no 4}  pp 437--456 (preprint math QA/9509008)

\bibitem{CIZ}    Cappelli A  Itzykson  C  and   Zuber J -B  1987  {The ADE
classification of minimal and $A_{1}^{(1)}$ conformal invariant theories }
{\it Commun  Math  Phys}  {\bf 13} p 1  

\bibitem{Coque:Qtetra}    Coquereaux R  2002  {Notes on the quantum
tetrahedron}  {\it Moscow Math  J } {\bf  2  no 1}   pp 1-40  (preprint hep-th/0011006)

\bibitem{GilCoque:ADE}    Coquereaux R  and Schieber  G 2002   {Twisted partition
functions for $ADE$ boundary conformal field theories and Ocneanu algebras of
quantum symmetries}  {\it  J  of Geom  and Phys}  {\bf 781}   pp 1-43  (preprint hep-th/0107001)

\bibitem{GilRobert2}   Coquereaux R  and  Schieber G 2003   {Determination of quantum
symmetries for higher $ADE$ systems from the modular $T$ matrix} 
{\it J  of Math  Phys}  {\bf 44}  pp  3809-3837 (preprint  hep-th/0203242)

\bibitem{CoqueTrinchero:cells}   Coquereaux R and   Trinchero R 2004 
 {On quantum symmetries of ADE graphs} 
{\it Adv in Theor  and Math   Phys}    {\bf 8   no 1}   (preprint hep-th/0401140)

\bibitem{RobertGilSL3Categories}    Coquereaux R and Schieber G 2007   { Orders and dimensions for sl(2) or sl(3) module categories and Boundary 
Conformal Field Theories on a torus}  {\it J  Math  Phys}  {\bf 48}  p 043511   (preprint  math-ph/0610073) 

\bibitem{DiFrancescoZuber}  Di Francesco P and   Zuber J -B  1990  {$SU(N)$-lattice integrable models associated with graphs}  {\it Nucl  Phys}  {\bf B338}  p 602  

\bibitem{YellowBook} Di Francesco  P Matthieu P  and  Senechal   D 1997   {\it Conformal Field Theory} (Berlin: Springer)

\bibitem{EtingofOstrik}   Etingof  P  and Ostrik  V  2004  {Finite tensor categories}  {\it Moscow Math J}   {\bf  4 no 3}   (preprint math QA/0301027)

\bibitem{FuchsRunkelSchweigert-I}   Fuchs J  Runkel  I Schweigert   C   2002  {TFT construction of RCFT correlators I: Partition functions}   {\it Nucl Phys}  {\bf B 646}  pp  353-497  (preprint hep-th/0204148)

\bibitem{Gannon-SU3} Gannon   T  1992   {The classification of
affine su(3) modular invariant partition functions}  {\it Commun  Math 
Phys}   {\bf 161}  pp 233--263  (preprint hep-th/9212060) 

\bibitem{Gannon-SU3-bis} Gannon   T  1992   {The classification of SU(3) modular invariants revisited}  {\it Annales de l'Institut  Poincar\'{e}} Phys  Theor {\bf 65}  pp  15-55 (preprint  hep-th/9404185)

\bibitem{PDO}   Goddard  P  Nahm W  and   Olive D 1985   {Symmetric spaces  Sugawara's energy momentum tensor in two dimensions and free fermions}  {\it Phys  Lett}  {\bf 160B} pp  111-116

\bibitem{GilDahmaneHassan}   Hammaoui D  Schieber  G  and Tahri  E H 2005   { Higher Coxeter graphs associated to affine $su(3)$ modular invariants}  {\it J  Phys} {\bf A38}  pp  8259-8286  (preprint hep-th/0412102)

\bibitem{EstebanGil}   Isasi E and   Schieber  G   2007 {From modular invariants to graphs: the modular splitting method}  
{\it J  of Physics A}  {\bf 40}  pp 6513-6537 (preprint math-ph/0609064)

\bibitem{Kac-Peter}   Kac V and   Peterson D  1984 {Infinite dimensional Lie algebras,  theta functions, and modular forms}  {\it Adv  Math}  {\bf 53}  p 125   

\bibitem{KacWakimoto} Kac V and  Wakimoto M 1988 {Modular and conformal invariance constraints in representation theory of affine algebras}   {\it Adv  Math} {\bf  70}  p 156  

\bibitem{NikshychVainerman}   Nikshych D and   Vainerman  2002 L {Finite quantum groupo\"\i ds and their applications}  in  New directions in Hopf algebras     {\it  Math  Sci  Res  Inst  Publ }  {\bf 43}  pp  211-262 Cambridge Univ  Press  Cambridge    (preprint mathQA/0006057) 

\bibitem{NikshychTuraevVainerman}  Nikshych D    Turaev  V and  Vainerman L  2003   {Invariant of  knots and 3-manifolds from quantum groupo\"\i ds}  {\it Topology and its Applications} {\bf 127  no  1-2 } pp 91-123  (preprint math QA/0006078)

\bibitem{Nill}   Nill F 1998  {Axioms for weak bialgebras}  (preprint mathQA/9805104)

\bibitem{Ocneanu:Unpublished}   Ocneanu  A 1996   {\it Unpublished}

\bibitem{Ocneanu:paths} Ocneanu  A 1999   {Paths on Coxeter diagrams: from 
Platonic solids and singularities to minimal models and subfactors}  Notes taken by Goto S 
{\it Fields Institute Monographs}  ed Rajarama Bhat et al  (AMS)   

\bibitem{Ocneanu:Bariloche}  Ocneanu  A   2000  {The Classification of  subgroups of quantum
SU(N)}  Lectures at Bariloche Summer School  Argentina   {\it  AMS Contemporary
Mathematics} {\bf  294}   ed  Coquereaux R  Garc\'{\i}a A and Trinchero R

\bibitem{Ostrik}   Ostrik  V 2003   {Module categories  weak Hopf algebras and modular invariants}  
{\it Transformation  groups}   {\bf  8 no 2}    pp 177-206 (preprint  math QA/0111139)  

\bibitem{PetkovaZuberNP1995}  Petkova V B and   Zuber J-B  1996   {From CFT to graphs}  {\it Nucl. Phys.} {\bf  B 463} pp 161-193 (preprint hep-th/9510175)

\bibitem{PetkovaZuberNP1997} Petkova V B and   Zuber J-B  1997 {Conformal field theory and graphs}  In Proceedings Goslar 1996  ÒGroup 21Ó   (preprint hep-th/9701103)

\bibitem{PetkovaZuber:Oc}   Petkova V B and   Zuber J-B  2001  { The many faces
of Ocneanu cells}  {\it Nucl  Phys}  {\bf B603}  pp 449-496   (preprint hep-th/0101151) 

\bibitem{SchellekensWarner}   Schellekens A N and   Warner  N P 1986   {Conformal subalgebras of Kac-Moody algebras} {\it  Phys Rev D}  {\bf 34 no 10}     pp  3092 -3096 

\bibitem{SchellekensYankielowicz} Schellekens A N  and Yankielowicz S 1990 {Modular invariants and fixed points} {\it Int J Mod Phys} {\bf A 5} 2903

\bibitem{Gil:Thesis}   Schieber G 2003    {L'alg\`ebre des sym\'etries quantiques d'Ocneanu et la classification des syst\`emes conformes \`a 2D}  PhD thesis in French or in Portuguese (Marseille: UP  and Rio de Janeiro: UFRJ) (preprint math-ph/0411077)    

\bibitem{Wassermann} Wassermann A   1998 {Operator algebras and conformal field theory  III  Fusion of positive energy representations of LSU(N) using bounded operators}  {\it Invent  Math}  {\bf 133}   pp 467-538 

\bibitem{FengXu}   Xu F  1998  {New braided endomorphisms from
conformal inclusions}  {\it Commun  Math  Phys}  {\bf  192 no 2} pp 349--403 


\end{thebibliography}
\end{document}